\documentclass[aps,prd,11pt,superscriptaddress,notitlepage,longbibliography,nofootinbib,tightenlines]{revtex4-1}

\usepackage{amsmath,amssymb,amsfonts}
\usepackage{graphicx}
\usepackage{subfigure}
\usepackage{dsfont}
\usepackage{slashed}

\usepackage{tikz}
\usepackage{color}

\usepackage{hyperref}
\definecolor{darkred}{rgb}{0.8,0.1,0.1}
\hypersetup{colorlinks=true, linkcolor=darkred, citecolor=blue, linktoc=page}

\newcommand{\N}[1]{\ensuremath{\mathcal N=#1}}

\DeclareMathOperator{\sech}{sech}
\DeclareMathOperator{\csch}{csch}
\DeclareMathOperator{\tr}{tr}

\newcommand{\GammaAdS}{\ensuremath{\Gamma_{\mathrm{AdS}}}}
\newcommand{\GammaS}{\ensuremath{\Gamma_{\mathrm{S}^5}}}
\newcommand{\RAdS}{\ensuremath{R_\mathrm{AdS}}}
\newcommand{\tRAdS}{\ensuremath{\tilde R_\mathrm{AdS}}}
\newcommand{\RS}{\ensuremath{R_{\mathrm{S}^5}}}
\newcommand{\tRS}{\ensuremath{\tilde R_{\mathrm{S}^5}}}

\newcommand{\GammaUalpha}{\ensuremath{\Gamma_{\underline{\vec{\alpha}}}}}
\newcommand{\GammaUbeta}{\ensuremath{\Gamma_{\underline{\vec{\beta}}}}}

\newcommand{\m}{\ensuremath{\mathfrak{m}}}
\newcommand{\n}{\ensuremath{\mathfrak{n}}}

\newcommand{\CO}{\ensuremath{\mathcal O}}
\newcommand{\eps}{\ensuremath{\epsilon}}

\makeatletter\def\l@subsubsection#1#2{}%
\makeatother

\begin{document}

\title{Supersymmetric D3/D5 for massive defects on curved space}

  \author{Brandon Robinson}
  \email{robinb22@uw.edu}
  \affiliation{Department of Physics, University of Washington, Seattle, WA 98195-1560, USA}
  
  \author{Christoph F.~Uhlemann}
  \email{uhlemann@physics.ucla.edu}
  \affiliation{Department of Physics, University of Washington, Seattle, WA 98195-1560, USA}
  \affiliation{Mani L.\ Bhaumik Institute for Theoretical Physics, Department of Physics and Astronomy, 
  University of California, Los Angeles, CA 90095, USA}

\begin{abstract}
We construct the holographic dual for \N{4} SYM on $\rm S^4$ and $\rm AdS_4$ coupled to massive \N{2} supersymmetric quenched flavor fields on a codimension-1 defect,
which is $\rm S^3$ and $\rm AdS_3$, respectively.
The holographic description is in terms of a D3/probe D5 brane system.
We set up and reduce the BPS equations for D5-brane embeddings with arbitrary supersymmetric deformations and partly solve them at the non-linear level.
The remaining equations are solved explicitly in a small-mass expansion.
We compute the contribution of the defect fields to the partition function on S$^4$ and compare to a field theory computation using supersymmetric localization,
for which we set up the matrix model.
Both computations agree, lending strong support to holographic probe brane constructions using D3/D5 configurations in general.
\end{abstract}

\maketitle
\tableofcontents

\baselineskip 14pt

\section{Introduction and Summary}

Quantum field theory on curved space has recently seen a revival of interest for purely formal reasons:
By reducing the path integral to a smaller subset of field configurations, supersymmetric localization allows one to exactly and non-perturbatively compute BPS observables such as the partition function or supersymmetric Wilson loops \cite{Pestun:2007rz}. To guarantee well-defined answers, compact Euclidean spaces are preferable, and to realize a maximal amount of symmetry spheres are a natural first choice.
As a simple generalization to backgrounds with a lesser amount of symmetry, 
continuous deformations such as squashed spheres have been studied \cite{Hama:2011ea}.

In particular in connection with the conjectured AdS/CFT dualities, the prospect of obtaining exact results at strong coupling offers the chance to confront the holographic dualities with decisive tests.
With full non-perturbative results from purely field-theoretic calculations in hand, one can compare to a holographic computation of the same quantity and test whether the two pictures agree.
The tests performed to date include dualities for field theories in various dimensions \cite{Bobev:2013cja, Marino:2011nm, Alday:2014rxa}, dualities involving probe brane constructions \cite{Karch:2015kfa} and many more.
While agreement of the two calculations is still not a formal proof for any of the dualities, these tests against localization calculations provide a powerful tool to potentially disprove the dualities, and the agreement on highly non-trivial observables provides strong circumstantial evidence for the validity of the dualities.

In interesting recent developments, first steps were taken to study localization on anti-de Sitter spacetimes \cite{David:2016onq, Bonetti:2016nma}.
In \cite{David:2016onq}, calculations on AdS$_2\times$S$^1$ were performed and checked, by relating the AdS$_2\times$S$^1$ geometries to covering spaces of S$^3$.
Studying field theories on AdS is interesting for a number of reasons, not the least of which is that it introduces a boundary and dependence on boundary conditions in a maximally symmetric way.
The study of boundary conditions on AdS has been pioneered by Breitenlohner and Freedman \cite{Breitenlohner:1982bm,Breitenlohner:1982jf}, and their results are ubiquitous in AdS/CFT. 
Their studies focus on free or perturbatively interacting fields, and 
to study the impact of boundary conditions on AdS for interacting field theories the AdS/CFT correspondences once again have been of great utility \cite{Aharony:2010ay}.
Coming back to testing the dualities using localization, the main obstacles that need to be overcome to extend the tests done so far to include AdS spacetimes are on the field theory side -- on the holographic side the calculations for field theories on AdS are closely parallel to the ones for spheres.
In fact, certain extra subtleties, e.g.\ with Euclidean supersymmetry, can even be avoided on AdS.
Holographic results can thus provide useful guidance for the field theory developments and facilitate consistency checks.
A study involving the impact of boundary conditions in a different context using localization was performed in \cite{Gava:2016oep}.

In both cases, for field theories on AdS$_d$ and on S$^d$, constructing the holographic dual amounts to finding asymptotically-AdS$_{d+1}$ supergravity solutions where the metric at the conformal boundary of AdS$_{d+1}$ is AdS$_d$ or S$^d$, respectively.
For a conformally invariant theory the supergravity solution can still be globally AdS$_{d+1}$, and the boundary metrics for AdS$_d$ and S$^d$ can be realized by a mere change of coordinates.
This reflects that AdS$_d$ and S$^d$ are representatives of the same (conformally flat) conformal structure as flat space.
However, to obtain physically meaningful results, it is often necessary to study deformations of the CFT that introduce a scale.
An example is the sphere partition function in even dimensions. 
It is a simple example of a computable quantity in localization and naturally suggests itself for comparison to holographic results. But it can be shifted by constants due to the presence of finite counterterms and is therefore not independent of the choice of renormalization scheme. 
This can be cured by studying supersymmetric mass deformations of the CFT of interest, 
such that the partition function becomes a function of the dimensionless product of mass parameter and radius of the sphere.
Once conformal invariance is broken, it does make a difference whether the theory is considered on flat space, on AdS$_d$ or on S$^d$,
and finding the holographic duals for the mass deformations becomes more involved.
But the free energy then contains scheme-independent and non-trivial information and allows for non-trivial tests.
This is the strategy followed for the tests on S$^4$ e.g.\ in \cite{Bobev:2013cja} and \cite{Karch:2015kfa}.

The results of \cite{Karch:2015kfa} showed that the field theory calculation of the free energy for SU($N_c$) \N{4} SYM coupled to $N_f\ll N_c$ massive hypermultiplets in the fundamental representation, at strong coupling on S$^4$, agrees with the holographic calculation of the same quantity.
The holographic computation starts out from the AdS$_5\times$S$^5$ solution to type IIB supergravity, into which $N_f$ probe D7 branes describing the flavor degrees of freedom are embedded.
For $N_f\ll N_c$ the backreaction of the D7 branes can be neglected, which considerably simplifies the computation:
Since the mass deformation only affects the flavor fields, the background \N{4} SYM theory can be realized on S$^4$ by a mere change of coordinates on AdS$_5$. Only the D7 branes are affected by the mass deformation and realizing the holographic setup amounts to finding the appropriate embedding into AdS$_5\times$S$^5$.
We found those embeddings for both, AdS$_4$ and S$^4$ in \cite{Karch:2015vra}.
The result for S$^4$ allowed us to test the holographic calculation against field theory results, where the probe limit $N_f \ll N_c$ indeed simplified the calculations drastically as well. 
The AdS$_4$ case led to interesting phenomenology which we studied in some detail in \cite{Karch:2015vra}.
The choice of boundary conditions at the conformal boundary of the AdS$_4$ field theory geometry indeed plays a crucial
role, and the free energy obtained from those embeddings should provide a good benchmark for field theory calculations.

\begin{figure}
\centering
\begin{tikzpicture}
 \node at (0,0) {\includegraphics[width=0.25\linewidth]{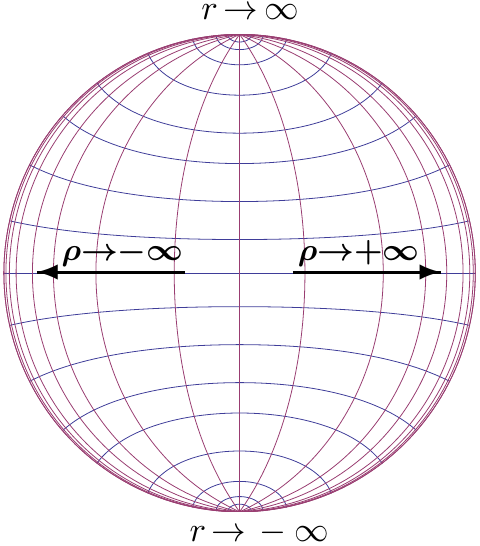}};
 \draw [very thick,dashed] (0,2.04) -- (0,-2.04);
 \draw[very thick, red] plot [smooth] coordinates { (0,2.04) (0.02,1.6) (0.3,0) (0.02,-1.6) (0,-2.04)};
\end{tikzpicture}
\caption{
Poincar\'e disc representation of the AdS$_4$ slicing of AdS$_5$ in coordinates (\ref{eqn:intro-coords1}), (\ref{eqn:intro-coords2}).
Thin vertical curves are $\rm AdS_4$ slices, while the horizontal curves correspond to slices of constant $r$.
The black dashed line shows the AdS$_4$ slice wrapped by the D5 brane for massless flavors, while the thick solid red curve schematically shows an embedding for massive flavors.
\label{fig:embedding}
}
\end{figure}

In this work we will further extend the holographic studies of field theories accessible to localization calculations.
We will stay within the realm of probe brane constructions, since they allow to add interesting features to the field theory with only a moderate amount of additional complexity, 
and consider a quantum field theory with a codimension-1 defect.
More specifically, we will consider the D3/D5 system \cite{Karch:2002sh, Aharony:2003qf,DeWolfe:2001pq},\footnote{%
A large body of work on D3/D5 systems based on integrability can be found in \cite{Buhl-Mortensen:2016jqo,Buhl-Mortensen:2016pxs,deLeeuw:2015hxa,deLeeuw:2017dkd,Widen:2017uwh},
a recent study of Wilson lines is in \cite{Preti:2017fhw} and a discussion of Mellin representations in dCFT can be found in \cite{Rastelli:2017ecj}.
}
describing SU($N_c$) \N{4} SYM coupled to a number $N_f\ll N_c$ of fundamental hypermultiplets confined to a three-dimensional subspace.
We will consider AdS$_4$ and S$^4$ backgrounds for the \N{4} SYM theory, with defects given by $\rm AdS_3$ and $\rm S^3$, respectively.
We will focus on $\rm AdS_3$ defects for the first part and discuss the $\rm S^3$ case afterwards using analytic continuation.
The defect breaks part of the isometries of the background as well as the conformal symmetries acting non-trivially in the directions transverse to the defect. 
But as long as no additional dimensionful parameters are introduced, the theory still preserves a combination of the broken isometries and conformal isometries, which combine with the transformations along the defect to the defect conformal symmetry SO($2,3$).
This is the same group for $\rm AdS_3$ defects and flat $\mathds{R}^{1,2}$ defects, and turns into $\rm SO(1,4)$ for $\rm S^3$. 
Correspondingly, already for a flat defect it turns out to be fruitful to think of the ``background'' \N{4} SYM as defined on two copies of AdS$_4$ joined at their respective conformal boundaries, which makes the preserved symmetries more manifest.
Holographically this again simply amounts to a change of coordinates on AdS$_5$ to realize an $\rm AdS_4$ slicing.
The supersymmetric embedding of D5 branes, described by the DBI action
\begin{align}
 S_{\mathrm{D5}}&=-T_5\int_{\Sigma_6}d^6\xi \sqrt{-\det\left(g+F\right)}+ T_5\int_{\Sigma_6}C_4\wedge F~,
\end{align}
in all cases with preserved defect conformal symmetry wraps an AdS$_4$ subspace of AdS$_5$ and an S$^2$ in S$^5$.
The embedding is illustrated in fig.~\ref{fig:embedding}.
But as soon as we wish to consider mass deformations or otherwise break conformal symmetry,
it makes a difference whether the defect is flat $\mathds{R}^{1,2}$, $\rm AdS_3$ or $\rm S^3$.
This amounts to the need for different D5-brane embeddings for $\rm AdS_3$ and $\rm S^3$ defects as compared to a flat $\mathds{R}^{1,2}$ defect (for which massive embeddings have been discussed already in \cite{Karch:2002sh}).

The D5 brane embeddings for massive flavors are expected to take the general form illustrated in fig.~\ref{fig:embedding}, 
and constructing them explicitly is the task at hand.
For the AdS$_5\times$S$^5$ background geometry created by the D3 branes we choose coordinates such that
\begin{align}
\label{eqn:intro-coords1}
 g_{\mathrm{AdS}_5\times\mathrm{S}^5}=d\rho^2+\cosh^2\!\rho\, g_{\mathrm{AdS}_4}
 +d\theta^2+\cos^2\!\theta\, g_{\mathrm{S}^2}+\sin^2\!\theta\, g_{\mathrm{\tilde S}^2}~,
\end{align}
with $\rho\in\mathds{R}$. 
For the $\rm AdS_4$ metric we choose an $\rm AdS_3$ slicing to realize an $\rm AdS_3$ defect.
We will need explicit parametrizations of the AdS$_4$ and S$^2$ metrics and choose
\begin{align}
\label{eqn:intro-coords2}
 g_{\mathrm{AdS}_4}&=dr^2+\cosh^2\! r\,g_{\mathrm{AdS}_3}~,
 &
 g_{\mathrm{S}^2}&=d\beta_1^2+\sin^2\beta_1 d\beta_2^2~.
\end{align}
The conformal boundary consists of the two $\rm AdS_4$ geometries obtained as $\rho\rightarrow\pm\infty$, which are joined at their respective conformal boundaries.
Moreover, sending $r\rightarrow\pm \infty$ for fixed $\rho$ yields two AdS$_3$ geometries, which are also joined at their conformal boundaries and correspond to a codimension-1 subspace in the boundary geometry.
For the D5-brane embedding we choose static gauge with $(r, \beta_1,\beta_2)$ and the AdS$_3$ directions as worldvolume coordinates,
such that the embedding is characterized by the slipping mode, $\theta$, parametrizing the S$^2$ wrapped inside S$^5$, 
and the bending mode, $\rho$, parametrizing the $\rm AdS_3$ slice wrapped in AdS$_5$. 
The slipping and bending modes are both restricted to be independent of the $\rm AdS_3$ coordinates.
The D5-branes intersect the conformal boundary at the two AdS$_3$ geometries obtained as $r\rightarrow\pm\infty$ and describe fundamental fields localized to this subspace of the conformal boundary.
The embedding generally preserves the $\rm AdS_3$ isometries, and for identically vanishing $\rho$ wraps an entire $\rm AdS_4$ slice.
Once the defect flavors are massive, defect conformal symmetry is broken and the subspace wrapped in AdS$_5$ is not necessarily AdS$_4$ anymore. 
Moreover, turning on the mass deformation preserves only a  U($1$) subgroup of the SU($2$) R-symmetry, and correspondingly the embedding only preserves a $\rm U(1)$ subgroup of the $\rm SU(2)$ isometries of the S$^2$ in S$^5$.
We can choose the coordinates such that the U($1$) acts by shifting the polar angle $\beta_2$, and consequently the worldvolume fields are independent of $\beta_2$.
That means the slipping and bending mode can both depend on the radial coordinate $r$ as well as on the azimuthal angle on the S$^2$, $\beta_1$, and the BPS equations that need to be solved to find supersymmetric embeddings are genuine PDEs.
In addition to that, we find that massive supersymmetric embeddings generically require -- in addition to non-trivial slipping and bending modes -- also flux on the S$^2$, i.e.\ a non-trivial worldvolume gauge field on the D5 branes.
In field theory terms, the requirement for non-trivial fields in addition to the slipping mode sourcing the mass term represents the need for an extra term required to accompany the mass deformation for supersymmetry to be preserved on curved space, as discussed systematically in \cite{Festuccia:2011ws}.
Finding supersymmetric embeddings thus amounts to solving non-linear coupled PDEs for three functions, $\rho$, $\theta$ and $A_{\beta_2}$, of two variables, $r$ and $\beta_1$.

Setting up the BPS equations for the D5 branes and reducing them to a minimal set of consistent equations constitutes a non-trivial technical part of the analysis, and we solve it in sec.~\ref{sec:non-linear-kappa}. 
We set up the conditions imposed by $\kappa$ symmetry \cite{Bergshoeff:1996tu, Cederwall:1996pv,Cederwall:1996ri} for our embedding ansatz spelled out in sec.~\ref{sec:ansatz}, and derive the eight resulting constraints on the embeddings.
The equations have non-linear coefficient functions and are non-linear in derivative terms as well. 
Nevertheless, we can isolate an equation that can be solved for the gauge field in terms of the slipping and bending modes.
The result is,
with $\tanh(r/2)=\tan(z/2)$, $x=\cos\beta_1$ and a parameter $\lambda$ satisfying $\lambda^2=1$,
\begin{align}
 A_{\beta_2}&=\lambda\tan z \cosh\rho\sin\theta-x\sinh\rho\cos\theta+\mathcal A_0~,
\end{align}
where $\mathcal A_0$ is a constant.
The remaining equations can then be reduced to a minimal set of two equations determining slipping and bending mode.
Together with the solution for the gauge field, they imply the entire set of conditions imposed by $\kappa$-symmetry.
The remaining two equations are
\begin{subequations}\label{eq:intro-bps}
\begin{align}
 -(1-x^2)\cosh\rho\, G_x+\sinh\rho\cos\theta F_z&=0~,
 \\
 \sinh\rho\,(G_z F_x-G_x F_z)-\cos^3\!\theta\, G_z -\sec^2\!z\, \cosh^3\!\rho\,\cos^3\!\theta&=0~.
\end{align}
\end{subequations}
with $G_{\mathfrak{a}}=\lambda\sinh\rho\sin^2\theta\,\partial_{\mathfrak{a}}(x\cot\theta)-\partial_{\mathfrak{a}}(\tan z \cosh\rho)$ and $\mathfrak{a}=r,x$.
At that level of generality, they encode not only mass deformations, but deformations by any of the supersymmetric operators sourced by combinations of slipping mode, bending mode and gauge field. More specifically, expanding all three fields in harmonics on S$^2$ yields a discrete series of fields on the AdS part of the background geometry. Each of these fields sources a different operator.
The remaining equations do not contain square roots, as would be typical for the equations of motion resulting from DBI actions with non-trivial Wess-Zumino terms, but we have not been able to solve them in closed form.
The equations can, however, be solved straightforwardly in a perturbative expansion around the massless embedding with no other sources or vacuum expectation values turned on, which we will refer to as zeroth-order embedding.

In sec.~\ref{sec:general-linearized} we study perturbative solutions for the embeddings. 
That is, we expand $\rho$, $\theta$ around the zeroth-order embedding and solve the equations (\ref{eq:intro-bps}). 
We work out the general linearized solution and find a discrete family of perturbations around the zeroth-order embedding. 
The source and vacuum expectation value for each supersymmetric combination of operators corresponding to the fluctuations
can be chosen independently on one of the $\rm AdS_3$ making up the defect.
On the remaining $\rm AdS_3$ the source and expectation value are then completely fixed.
In particular, even with no sources turned on we find a ``moduli space'' of supersymmetric states.
This is reminiscent of the family of supersymmetric states parametrized by the chiral condensate found for D3/D7 in \cite{Karch:2015vra}.
At the technical level, the reason for the option to dial source and one-point function independently in both cases is that there is no constraint from 
the requirement for regularity at the $r=0$ slice.
Physically, the different states are expected to correspond to different choices of boundary conditions for the flavor fields, at the conformal boundary of AdS.
That is, the conformal boundary of $\rm AdS_4$ in the analysis of \cite{Karch:2015vra}, and the boundary of $\rm AdS_3$ for the defect flavors studied here.
We have given evidence for that interpretation in \cite{Karch:2015vra} which is suggestive but not conclusive.
In contrast to the analysis for D3/D7, where we restricted to a more special ansatz from the outset, we find a larger family of states for D3/D5 in the current analysis. 
Presumably, a similarly rich family would be found for D3/D7 with a more general ansatz.
For both cases, this offers interesting prospects for mutual benefits with localization calculations:
To match the free energy exactly, the choice of boundary conditions corresponding to each of our embeddings has to be matched precisely. The holographic calculations can give useful intuition for what results to expect from localization calculations, and the localization results in turn could help pin down precisely which boundary conditions correspond to which embedding.
We also discuss the embeddings to higher orders in the perturbative expansion, which gives some insight into the structure of the non-linear solutions.

In sec.~\ref{sec:S4-holographic} we analytically continue the BPS equations and embeddings for the AdS$_4$ slicing with $\rm AdS_3$ defect to global Euclidean AdS$_5$,
to describe \N{4} SYM on $\rm S^4$, coupled to massive defect fields on an equatorial $\rm S^3$.
The metric on the $\rm AdS_5$ part of the background geometry becomes
\begin{align}
 g_{\mathrm{AdS}_5}&=dR^2+\sinh^2\!R\Big[d\chi^2+\cos^2\!\chi\, g_{\mathrm{S}^3}\Big]~,
\end{align}
and the embedding of the D5 branes is characterized by a bending mode $\chi$, the slipping mode $\theta$ and the worldvolume gauge field as functions of $R$ and $\beta_1$.
The defect where the D5-branes add fundamental fields is the $\rm S^3$ obtained as $R\rightarrow\infty$, and the massless embedding corresponds to identically vanishing $\chi$, $\theta$ and gauge field.
We focus on a perturbative mass deformation, which corresponds to a non-trivial profile for the slipping mode.
The BPS equations then require non-trivial profiles for the bending mode and gauge field as well, 
and they turn out to be purely imaginary for a real mass deformation.
In field theory terms this reflects the fact that supersymmetric theories on $\rm S^4$ (or by analytic continuation on $\rm dS_4$) are in general not unitary, unless they happen to be conformally invariant as well.
More specifically, the coefficient of the supersymmetry-restoring extra term that has to accompany the mass deformation on $\rm AdS_4$ turns imaginary upon analytic continuation to $\rm dS_4$ or $\rm S^4$.
We found a similar phenomenon for the D3/D7 case studied in \cite{Karch:2015vra}, where the gauge field turned imaginary upon analytic continuation to $\rm dS_4$.
Unlike for the AdS$_4$ slicing, the regularity conditions at the origin of AdS$_5$ in the $\rm S^4$ slicing do fix the subleading terms of the D5-brane fields in terms of the leading terms,
and the vacuum state is unique.
We compute the one-point functions of the operators sourced by the supersymmetric combination of slipping mode, bending mode and worldvolume gauge field,
and also the contribution of the defect fields to the partition function on $\rm S^4$ to quadratic order in the mass deformation. 
The result is, with the identification of holographic and field theory parameters as summarized e.g.\ in \cite{Chang:2013mca},
\begin{align}\label{eq:deltaF-intro}
 \delta\mathcal F(\rm S^4)&=- \mu N_f N_c\left[\frac{4}{3} + \frac{2M^2}{\mu^2}+\dots\right]~,
\end{align}
where $\mu=\sqrt{\lambda}/2\pi$ with $\lambda$ the 't Hooft coupling and $M$ is the mass of the defect fields.
The dots denote subleading terms in the strong-coupling expansion and terms with higher powers of $M$.
Although the partition function of \N{4} SYM on $\rm S^4$ is scheme dependent,
we will argue that the terms at $\mathcal O(m^0)$ and $\mathcal O(m^2)$ in the defect contribution to the partition function are renormalization scheme independent 
and therefore physically meaningful quantities, that can be compared reasonably to a localization computation.

In sec.~\ref{sec:localization} we switch to the QFT side and compute the contribution of the defect fields to the partition function using supersymmetric localization.
We are not aware of a detailed discussion of localization in theories with defects, but the matrix model can be constructed from a combination of the results obtained 
for four-dimensional \N{4} SYM and independent localization computations in intrinsically three-dimensional Chern-Simons-matter theories. 
We discuss the derivation of the matrix model, for which we find
\begin{align}\label{eq:mm-intro}
 \mathcal Z_{\rm defect}&=
 \int da^{N_c-1}\prod_{i<j} a_{[ij]}^2\,\frac{1}{\prod_i \cosh^{N_f}\big(\pi(a_i + M)\big)}~e^{S_0}~,
 &
 S_0&={-\frac{8\pi^2}{\lambda}N_c\sum_i a_i^2}~.
\end{align}
From this matrix model we again compute the contribution of the defect fields to the partition function at strong coupling and in the quenched approximation.
The result precisely matches the holographic computation resulting in (\ref{eq:deltaF-intro}).
This lends strong support to the holographic computations using D3/D5, here and generally,
and also to the construction of the matrix model resulting in (\ref{eq:mm-intro}).

\section{Reduced BPS equations for \texorpdfstring{$\rm AdS_3$}{AdS3} defects}
In this section we will derive the BPS equations for general D5-brane embeddings into 
AdS$_5\times$S$^5$ with two copies of $\rm AdS_3$-sliced $\rm AdS_4$ as boundary geometry, which are compatible with the supersymmetries preserved by a mass deformation.
We are thus looking for configurations that preserve one quarter of the original 32 supersymmetries of AdS$_5\times$S$^5$:
adding the D5-brane describing the defect fields breaks the 32 supersymmetries of the $\rm AdS_5\times S^5$ solution down to 16,
and adding the mass deformation further breaks those down to 8 remaining supersymmetries.
What precisely the supercharges preserved by a given embedding are is dictated by $\kappa$-symmetry \cite{Bergshoeff:1996tu, Cederwall:1996pv,Cederwall:1996ri}.
Finding the BPS equations can be split up into two steps.
First, after setting up the embedding ansatz in sec.~\ref{sec:ansatz}, 
we will study embeddings with an infinitesimally small mass parameter to determine precisely which of the supersymmetries are preserved in sec.~\ref{sec:linearized-kappa}.
With that information in hand, we will then study the full non-linear $\kappa$-symmetry constraints and derive the
conditions for general embeddings to preserve the previously identified supersymmetries in sec.~\ref{sec:non-linear-kappa}.

We find a total of eight equations for the three functions $\rho$, $\theta$, $A_{\beta_2}$, 
which are all first-order PDEs but not independent.
We will isolate an equation which can be solved for the gauge field in sec.~\ref{sec:solving-A}
and show that the remaining equations can be reduced to two independent ones in sec.~\ref{sec:remainingBPS}.

\subsection{Background geometry and embedding ansatz}\label{sec:ansatz}

We want to study \N{4} SYM on two copies of AdS$_4$, with a codimension-1 defect localized to an AdS$_3$ slice.
To realize this geometry on the boundary of AdS$_5$, we choose coordinates where AdS$_5$ is covered by AdS$_4$ slices,
which are in turn covered by AdS$_3$ slices.
For the S$^5$ part it is convenient to make the S$^2$ which the D5 branes wrap explicit.
We therefore fix coordinates for the AdS$_5\times$S$^5$ background s.t.\
\begin{subequations}\label{eqn:metric-AdS5}
\begin{align}
 g_{\mathrm{AdS}_5\times\mathrm{S}^5}=d\rho^2+\cosh^2\!\rho\, g_{\mathrm{AdS}_4}
 +d\theta^2+\cos^2\!\theta\, g_{\mathrm{S}^2}+\sin^2\!\theta\, g_{\mathrm{\tilde S}^2}~.
\end{align}
We introduce explicit coordinates for AdS$_4$ and the two 2-spheres s.t.\
\begin{align}
 g_{\mathrm{AdS}_4}&=dr^2+\cosh^2\! r\,g_{\mathrm{AdS}_3}~,
 &
 g_{\mathrm{S}^2}&=d\beta_1^2+\sin^2\!\beta_1\, d\beta_2^2~,
 \\
 g_{\mathrm{AdS}_3}&=dx^2+e^{2x}(-dt^2+dy^2)~,
 &
 g_{\mathrm{\tilde S}^2}&=d\alpha_1^2+\sin^2\!\alpha_1\, d\alpha_2^2~.
\end{align}
\end{subequations}
In the probe limit the D5 branes are simply described by a DBI action with a Wess-Zumino term.
With $2\pi\alpha^\prime=1$ it takes the form\footnote{%
The sign convention for the WZ term is that of e.g.\ \cite{Cederwall:1996pv,Bergshoeff:1996tu,Kruczenski:2003be} and \cite{Karch:2015vra},
and differs from the choice in \cite{DeWolfe:2001pq}.
}
\begin{align}\label{eqn:D5-action}
 S_{\mathrm{D5}}&=-T_5\int_{\Sigma_6}d^6\xi \sqrt{-\det\left(g+F\right)}+ T_5\int_{\Sigma_6}C_4\wedge F~,
\end{align}
where $g$ is the pullback of the background metric. 
For the 4-form gauge field we take\footnote{%
This is the normalization for gauge field and the WZ term of \cite{DeWolfe:2001pq,Kruczenski:2003be}, where $dC_4=4L^{-1}\operatorname{vol}(\mathrm{AdS}_5)$.
It is different from the normalization used in \cite{Karch:2015vra}.
}
\begin{align}\label{eqn:RR-gauge-field}
 C_4&=L^{-1}\zeta(\rho)\operatorname{vol}(\mathrm{AdS}_4)+\dots~,&\zeta^\prime(\rho)=4\cosh^4\rho~.
\end{align}
We want embeddings preserving the full AdS$_3$ isometries and fix static gauge, using $(r,x,t,y,\vec{\beta})$ as coordinates on the D5 brane.
The embedding is then described by the slipping and bending modes
\begin{align}
 \theta&=\theta(r,\vec{\beta})~,
 &
 \rho=\rho(r,\vec{\beta})~,
\end{align}
respectively.
The general form of the gauge field compatible with preserving the AdS$_3$ isometries 
(after imposing radial gauge with $A_r=0$) is
\begin{align}
 A&=A_{\beta_1}(r,\vec{\beta})d\beta_1+A_{\beta_2}(r,\vec{\beta})d\beta_2~.
\end{align}
We expect supersymmetric embeddings to preserve a U($1$), which can be used to eliminate dependence on $\beta_2$, as described above.
For now, however, we will keep the dependence general.

\subsubsection{Background Killing spinors}
For the background Killing spinors we use the $\rm AdS_5\times S^5$ Killing spinor equation 
in the conventions of \cite{Grisaru:2000zn}, namely
\begin{align}\label{eqn:Killing-spinor-eq}
 D_\mu\epsilon&=\frac{i}{2}\,\GammaAdS\Gamma_\mu\epsilon~,\quad \mu=0\dots 4~,&
 D_\mu\epsilon&=\frac{i}{2}\,\GammaS\Gamma_\mu\epsilon~,\quad \mu=5\dots 9~.&
\end{align}
Whenever explicit values for indices appear, we will use an underline to distinguish local Lorentz indices from coordinate indices.
We then have $\GammaAdS=\Gamma^{\underline{\rho r xty}}=-\Gamma_{\underline{\rho r xty}}$ and $\GammaS=\Gamma^{\underline{\theta \alpha_1\alpha_2\beta_1\beta_2}}$.
The Killing spinors are given by
\begin{align}\label{eqn:Killing-R}
 \epsilon&=\RAdS\RS \epsilon_0~,
\end{align}
where the AdS$_5$ part of the $R$-matrix, with $P_{x\pm}=\frac{1}{2}(\mathds{1}\pm i\Gamma_{\underline{x}}\GammaAdS)$, is
\begin{align}\label{eqn:AdS-R-matrix}
 \RAdS&=e^{\frac{i\rho}{2}\Gamma_{\underline{\rho}}\GammaAdS}e^{\frac{i r}{2}\Gamma_{\underline{r}}\GammaAdS} R_{\mathrm{AdS}_3}~,
 &
 R_{\mathrm{AdS}_3}&=e^{\frac{i x}{2}\Gamma_{\underline{x}}\GammaAdS}
 +ie^{\frac{x}{2}}\left(t\Gamma_{\underline{t}}+y\Gamma_{\underline{y}}\right)\GammaAdS P_{x-}~.
\end{align}
The S$^5$ part, with $\GammaUalpha=\Gamma_{\underline{\alpha_1}}\Gamma_{\underline{\alpha_2}}$ and analogously for $\GammaUbeta$, reads
\begin{align}\label{eq:R-matrices}
 \RS&=e^{\frac{i\theta}{2}\Gamma_{\underline{\theta}}\GammaS}R_{\mathrm{\tilde S}^2} R_{\mathrm{S}^2}~,
 &
 R_{\mathrm{\tilde S}^2}&=e^{\frac{\alpha_1}{2}\Gamma_{\underline{\theta}}\Gamma_{\underline{\alpha_1}}} e^{\frac{\alpha_2}{2}\GammaUalpha}~,
 &
 R_{\mathrm{S}^2}&=e^{\frac{i\beta_1}{2}\Gamma_{\underline{\beta_1}}\GammaS}
 e^{\frac{\beta_2}{2}\GammaUbeta}~.
\end{align}
This completes the discussion of the Killing spinors. 
For later convenience we define charge-conjugated $R$-matrices with a tilde as $\tRAdS=C(\RAdS)^\star C$ and analogously for the other $R$-matrices.
Some useful identities are then
\begin{align}\label{eqn:Rtilde-identities}
 \tRAdS&=e^{-i\rho\Gamma_{\underline{\rho}}\GammaAdS}\Gamma_{\underline{\rho}}\RAdS\Gamma_{\underline{\rho}}~,
 &
 \tRS&=-e^{-i\theta\Gamma_{\underline{\theta}}\GammaS}\GammaUbeta\RS\GammaUbeta~.
\end{align}

\subsection{\texorpdfstring{$\kappa$}{kappa}-symmetry generalities}
The $\kappa$-symmetry condition as spelled out e.g.\ in \cite{Bergshoeff:1996tu} is a projection condition on the background Killing spinors of the form $\Gamma_\kappa\epsilon=\epsilon$. 
For an embedding to preserve some supersymmetry, the condition needs to have non-trivial solutions, and this provides the constraints which are the BPS equations.
For the ansatz spelled out above
\begin{align}
 \Gamma_\kappa&=\frac{1}{\sqrt{\det(1+X)}}\left(J_{(5)}^{(0)}+\frac{1}{2}\gamma^{jk}F_{jk}J_{(5)}^{(1)}\right)~,
 &
 J_{(5)}^{(n)}&=(-1)^n(\sigma_3)^{n+1}i\sigma_2\otimes \Gamma_{(0)}~.
\end{align}
Since the field strength $F$ only has non-trivial components in the $r,\vec{\beta}$ directions, the usual sum in $\Gamma_\kappa$
terminates after the linear term.
As in \cite{Karch:2015vra}, we switch to complex notation, such that
\begin{align}
 J_{(5)}^{(0)}\begin{pmatrix}\epsilon_1\\ \epsilon_2\end{pmatrix}&=i C\left(\Gamma_{(0)}\epsilon\right)^\star~,
 &
 J_{(5)}^{(1)}\begin{pmatrix}\epsilon_1\\ \epsilon_2\end{pmatrix}&=i\Gamma_{(0)}\epsilon~.
\end{align}
The $\kappa$-symmetry condition then becomes
\begin{align}
 iC\left(\Gamma_{(0)}\epsilon\right)^\star+\frac{i}{2}\gamma^{ij}F_{ij}\Gamma_{(0)}\epsilon&=\sqrt{\det(1+X)}\epsilon~,
 &
 \Gamma_{(0)}&=\frac{1}{6!\sqrt{-\det g}}\varepsilon^{i_1\dots i_6}\gamma_{i_1\dots i_6}~,
\end{align}
where $\gamma_i=e_i^a\Gamma_a$ and $e^a=E_\mu^a(\partial_i X^\mu)dx^i$ is the pullback of the ten-dimensional
vielbein $E^a$ to the D5 worldvolume.
To evaluate $\Gamma_{(0)}$ we need $e^a$.
The straightforward part is
\begin{align}
 e^{\underline{\alpha_1}}=e^{\underline{\alpha_2}}&=0~,
 &
 e^a&=E^a~,\quad
 a=\underline{\beta_1}, \underline{\beta_2}, \underline{t}, \underline{y}, \underline{x}, \underline{r}~.
\end{align}
For notational convenience we
introduce fraktur indices $\m,\n,\dots$ running over $r,\beta_1,\beta_2$,
and the remaining part then reads
\begin{align}
 e^{\underline{\theta}}&=d\theta=(\partial_{\m}\theta)d\xi^\m~,
 &
 e^{\underline{\rho}}&=d\rho=(\partial_{\m}\rho)d\xi^\m~.
\end{align}
With the explicit form of the pullback of the vielbein, $e^a$, we then find
\begin{align}
 \Gamma_{(0)}&=\frac{1}{\sqrt{-\det g}}\gamma_{rxty\beta_1\beta_2}
 =\frac{\cosh^3\!\rho\,\cosh^3\!r\,\sqrt{-\det g_{\mathrm{AdS}_3}}}{\sqrt{-\det g}}
 \,\hat \Gamma~,
\end{align}
where $g_{\mathrm{AdS_3}}$ is once again the metric on AdS$_3$ of unit curvature radius,
and $\hat\Gamma$ is given by
\begin{align}\label{eqn:Gammahat}
\hat\Gamma&=\Gamma_{\mathrm{AdS}3}\gamma_{r\beta_1\beta_2}~,
&
\Gamma_{\mathrm{AdS}3}&=\Gamma^{\underline{xty}}=-\Gamma_{\underline{xty}}~.
\end{align}
The explicit expressions for the involved $\gamma$-matrices are
\begin{align}\label{eqn:gammarbeta}
 \gamma_{\m}&=\Gamma_{\m}+(\partial_{\m}\rho)\Gamma_{\rho}+(\partial_{\m}\theta)\Gamma_{\theta}~.
\end{align}
Note that the $\Gamma$-matrices involve the (diagonal) AdS$_5\times$S$^5$ vielbein.
The complete $\kappa$-symmetry condition then becomes
\begin{align}
 iC\Big(\hat\Gamma\epsilon\Big)^\star+\frac{i}{2}\gamma^{ij}F_{ij}\hat\Gamma\epsilon&=h\epsilon~,
 &
 h&=\frac{\sqrt{-\det(g+F)}}{\cosh^3\!\rho\,\cosh^3\!r\,\sqrt{-\det g_{\mathrm{AdS}_3}}}~.
\end{align}
This condition, together with the Killing spinors given previously, provides the constraints for supersymmetric embeddings and will have to be evaluated more explicitly.
There are no explicit factors of $i$ in $\hat\Gamma$, so we can use $C^2=\mathds{1}$ and $C(\Gamma^\mu)^\star C=\Gamma^\mu$ to rewrite it.
Evaluating also $h$ more explicitly yields the for now final form of the condition
\begin{align}\label{eqn:kappa-condition}
 i \hat\Gamma C\epsilon^\star+\frac{i}{2}\gamma^{ij}F_{ij}\hat\Gamma\epsilon&=h\epsilon~,
 &
 h&=\sqrt{\det (g_{\m\n}+F_{\m\n})}~,
\end{align}
where
\begin{align}\label{eqn:h-M123}
 g_{\m\n}&=(g_{\mathrm{AdS}_5\times\mathrm{S}^5})_{\m\n}
 +
 (\partial_{\m}\rho)(\partial_{\n}\rho)+(\partial_{m}\theta)(\partial_{\n}\theta)~.
\end{align}

\subsection{Preserved supersymmetries}\label{sec:linearized-kappa}
In this section we determine the precise supersymmetries that are preserved by embeddings describing massive defect fields.
Since the form of the supersymmetries is independent of the value of the mass parameter, we can conveniently work with an infinitesimally small mass.

We first need to determine the symmetries preserved by the massless embedding with $\rho=A=\theta=0$. 
Denoting the order in the mass deformation by a superscript in brackets, we have
\begin{align}
 \hat\Gamma^{(0)}&=-h^{(0)} \Gamma_{\underline{\rho}}\GammaAdS\GammaUbeta~,
 &
 h^{(0)}&=\sin\beta_1~.
\end{align}
Note that $\GammaAdS=\Gamma^{\underline{01234}}=-\Gamma_{\underline{01234}}$.
For the massless embedding the $\kappa$-symmetry condition (\ref{eqn:kappa-condition}) becomes
\begin{align}
 i\hat\Gamma^{(0)} C{\epsilon^{(0)}}^\star&=h^{(0)}\epsilon^{(0)}
 &\Longleftrightarrow&&
 -i \Gamma_{\underline{\rho}}\GammaAdS\GammaUbeta C{\epsilon^{(0)}}^\star&=\epsilon^{(0)}~.
\end{align}
With $\tRAdS$ and $\tRS$ defined below (\ref{eq:R-matrices}), 
the resulting projection condition on the constant spinor $\epsilon_0$
in (\ref{eqn:Killing-R}) becomes
\begin{align}
 -i\RAdS^{-1}\RS^{-1}\Gamma_{\underline{\rho}}\GammaAdS\GammaUbeta\tRAdS\tRS C\epsilon_0^\star&=\epsilon_0~.
\end{align}
With (\ref{eqn:Rtilde-identities}) and $\rho=\theta=0$, this straightforwardly evaluates to
\begin{align}\label{eqn:massless-projector}
 -i\Gamma_{\underline{\rho}}\GammaAdS \GammaUbeta\,C\epsilon_0^\star&=\epsilon_0~.
\end{align}
This condition reduces the number of preserved supersymmetries from 32 to 16 and singles out precisely which supersymmetries are preserved by the massless embedding.

To find infinitesimally massive embeddings, we solve the $\kappa$-symmetry condition (\ref{eqn:kappa-condition}) at linear order in a small-fluctuation expansion around the massless embedding.
We set $\theta=\theta(r)$, i.e.\ assume that there is no dependence on the $\rm S^2$ coordinates, and assume $\theta$ to be small. 
Similar expansions are used for $\rho$ and $A$, except for that we do not constrain their dependence on $\rm S^2$.
Denoting by a superscript the order in the small-fluctuation expansion, the $\kappa$-symmetry condition at linear order then reads
\begin{align}
\label{eq:lin-kappa-1}
 i\hat\Gamma^{(1)} C{\epsilon^{(0)}}^\star+i\hat\Gamma^{(0)}C{\epsilon^{(1)}}^\star
 +\frac{i}{2}\gamma^{(0)ij}F^{(1)}_{ij}\hat\Gamma^{(0)}\epsilon^{(0)}
 &=h^{(0)}\epsilon^{(1)}~.
\end{align}
On the right hand side we have used that corrections to $h$ are at least quadratic, so $h^{(1)}=0$.
The Killing spinor at linearized order straightforwardly evaluates to
\begin{align}
 \epsilon^{(1)}&=\frac{i}{2}\left(\rho\Gamma_{\underline{\rho}}\GammaAdS+\theta\Gamma_{\underline{\theta}}\GammaS\right)\epsilon^{(0)}
 =:\delta R\,\epsilon^{(0)}~.
\end{align}
Noting that $C(\delta R)^\star C=-\delta R$, the $\kappa$-symmetry condition (\ref{eq:lin-kappa-1}) becomes
\begin{align}
 i\left(\hat\Gamma^{(1)} -\hat\Gamma^{(0)}\delta R\right)C{\epsilon^{(0)}}^\star
 +\frac{i}{2}\gamma^{(0)ij}F^{(1)}_{ij}\hat\Gamma^{(0)}\epsilon^{(0)}
 &=h^{(0)}\delta R\,\epsilon^{(0)}~.
\end{align}
Using $(\hat\Gamma^{(0)})^2={h^{(0)}}^2\mathds{1}$, the massless projection condition can be written as
$C{\epsilon^{(0)}}^\star=-i\hat\Gamma^{(0)}\epsilon^{(0)}/h^{(0)}$.
So we can eliminate $C{\epsilon^{(0)}}^\star$ and the $\kappa$-symmetry condition becomes
\begin{align}
 \frac{1}{h^{(0)}}\left(\hat\Gamma^{(1)} -\hat\Gamma^{(0)}\delta R\right)\hat\Gamma^{(0)}\epsilon^{(0)}
 +\frac{i}{2}\gamma^{(0)ij}F^{(1)}_{ij}\hat\Gamma^{(0)}\epsilon^{(0)}
 &=h^{(0)}\delta R\,\epsilon^{(0)}~.
\end{align}
We can now use $\hat\Gamma^{(0)}\delta R\,\hat\Gamma^{(0)}={h^{(0)}}^2\delta R$ to find
\begin{align}\label{eqn:kappa-lin-1}
 \frac{1}{h^{(0)}}\hat\Gamma^{(1)}\hat\Gamma^{(0)}\epsilon^{(0)}
 +\frac{i}{2}\gamma^{(0)ij}F^{(1)}_{ij}\hat\Gamma^{(0)}\epsilon^{(0)}
 &=2h^{(0)}\delta R\,\epsilon^{(0)}~.
\end{align}
For the computation of the term involving the field strength it is useful to note that $\gamma^{(0)ij} F_{ij}=\Gamma^{ij}F_{ij}$,
since for the zeroth-order embedding with $\rho=\theta=0$ the induced metric $g$ coincides with the corresponding part of $g_{\mathrm{AdS}_5\times\mathrm{S}^5}$ 
for the components of interest.
We will also set 
\begin{align}
\label{eq:A-lin-split}
A_{\beta_i}&=f(r)\omega_i(\vec{\beta})~, 
\end{align}
such that $A=f\omega$ with $\omega$ a one-form on S$^2$.
For this subsection we use the shorthand notation $\partial_{\beta_{1/2}}=\partial_{1/2}$ and a prime to denote derivatives w.r.t.\ $r$.
Using once again that $\rho=\theta=0$ for the zeroth-order embedding, we then find
\begin{align}
 \frac{1}{2}\gamma^{(0)ij}F_{ij}^{(1)}&=
 f^\prime\Gamma_{\underline{r}}\left(\omega_1\Gamma_{\underline{\beta_1}}+\omega_2\csc\beta_1\Gamma_{\underline{\beta_2}}\right)
 +f (\star d\omega)\GammaUbeta~,
\end{align}
where $\star d\omega=\csc\beta_1(\partial_1\omega_2-\partial_2\omega_1)$.
For the combination which appears in (\ref{eqn:kappa-lin-1}) this yields
\begin{align}
 \frac{i}{2}\gamma^{(0)ij}F_{ij}^{(1)}\hat\Gamma^{(0)}&=
 i\left[
 f^\prime\Gamma_{\underline{r}}\left(\omega_2\Gamma_{\underline{\beta_1}}-\sin\beta_1\omega_1\Gamma_{\underline{\beta_2}}\right)
 +
 f\sin\beta_1(\star d\omega)\,\mathds{1}
 \right]\Gamma_{\underline{\rho}}\GammaAdS~.
\end{align}
The last object we have to work out explicitly is $\hat\Gamma^{(1)}$ with $\hat\Gamma$ given in (\ref{eqn:Gammahat}).
The combination appearing in (\ref{eqn:kappa-lin-1}) becomes
\begin{align}
 \frac{1}{h^{(0)}}\hat\Gamma^{(1)}\hat\Gamma^{(0)}&=
 \sin\beta_1\left(\theta^\prime\Gamma_{\underline{\theta}}+\rho^\prime\Gamma_{\underline{\rho}}\right)\Gamma_{\underline{r}}
 +\sin\beta_1(\partial_1\rho)\Gamma_{\underline{\rho}}\Gamma_{\underline{\beta_1}}
 +(\partial_2\rho)\Gamma_{\underline{\rho}}\Gamma_{\underline{\beta_2}}~.
\end{align}
The $\kappa$-symmetry condition in eq.~(\ref{eqn:kappa-lin-1}), after dividing by $\sin\beta_1$, then takes the explicit form
\begin{align}
\begin{split}
 0=\Big[&
 \theta^\prime\Gamma_{\underline{\theta}}\Gamma_{\underline{r}}+\rho^\prime\Gamma_{\underline{\rho}}\Gamma_{\underline{r}}
 +(\partial_1\rho)\Gamma_{\underline{\rho}}\Gamma_{\underline{\beta_1}}
 +\csc\beta_1(\partial_2\rho)\Gamma_{\underline{\rho}}\Gamma_{\underline{\beta_2}}
 -i\rho\Gamma_{\underline{\rho}}\GammaAdS-i\theta\Gamma_{\underline{\theta}}\GammaS
 \\
 &
 -if^\prime\big(\omega_1\Gamma_{\underline{\beta_2}}-\omega_2\csc\beta_1\Gamma_{\underline{\beta_1}}\big)\Gamma_{\underline{\rho}}\Gamma_{\underline{r}}\GammaAdS
 +if(\star d\omega)\Gamma_{\underline{\rho}}\GammaAdS
 \Big]\epsilon^{(0)}~.
\end{split}
\end{align}
Multiplying by $\Gamma_{\underline{\theta}}\Gamma_{\underline{\rho}}$ and using that $\epsilon_0$ is a chiral spinor with 
$\Gamma_{11}\epsilon_0=\epsilon_0\Leftrightarrow\GammaS\epsilon=-\GammaAdS\epsilon$, yields
\begin{align}\label{eqn:lin-kappa-2}
\begin{split}
 \Big[\theta^\prime\Gamma_{\underline{\rho}}\Gamma_{\underline{r}}+i\theta\Gamma_{\underline{\rho}}\GammaAdS\Big]\epsilon^{(0)}
 &=
 \Big[
 i\big(\rho-f\star\!d\omega\big)\GammaAdS-(\partial_1\rho)\Gamma_{\underline{\beta_1}}-\csc\beta_1(\partial_2\rho)\Gamma_{\underline{\beta_2}}
 \Big]\Gamma_{\underline{\theta}}\epsilon^{(0)}
 \\&
 \hphantom{=}\ +\Big[
 i f^\prime\big(\omega_2\csc\beta_1\Gamma_{\underline{\beta_1}}-\omega_1\Gamma_{\underline{\beta_2}}\big)\Gamma_{\underline{r}}\GammaAdS
 -\rho^\prime\Gamma_{\underline{r}}
 \Big]\Gamma_{\underline{\theta}}\epsilon^{(0)}~.
\end{split}
\end{align}
The term on the left hand side has no non-trivial dependence on the S$^5$ coordinates. 
That is, after multiplying both sides of the equation by $\RS^{-1}$ all dependence drops out.
The first term on the right hand side only has dependence on the AdS$_5$ directions through $\rho$ and $f$,
no non-trivial AdS$_5$ $\Gamma$-matrix structures.
That means it is independent of the AdS$_3$ directions $x,t,y$ after multiplying both sides of the equation by $\RAdS^{-1}$.
The interesting term is the second one on the right hand side, in line two.
It has non-trivial dependence on the S$^5$ {\it and} on the AdS$_3$ directions, due to the 
appearance of $\Gamma_{\underline{r}}$.\footnote{%
At $\rho=0$, $\RAdS^{-1}\Gamma_{\underline{r}}\RAdS$ has four independent $\Gamma$-matrix structures whose coefficients
depend on $(x,t,y)$ in such a way that the dependence can not be cancelled by imposing a projector on $\epsilon_0$. 
See (\ref{eqn:calR-identities}) below.
}
That means it either has to vanish by itself, or at least one of the non-trivial dependences on the S$^5$ and the AdS$_3$ directions has to drop out
for it to cancel with one of the other terms in the equation.
But it is not possible to cancel the non-trivial AdS$_3$ dependence, so it must be the dependence on the S$^5$ directions which cancels.
This allows it to combine with the term on the left hand side, which also has non-trivial dependence on the AdS$_3$ directions.
In fact, all of the $\Gamma$-matrix structures in $\RAdS^{-1}\Gamma_{\underline{r}}\RAdS$ have non-trivial dependence 
on the AdS$_3$ directions. So in the first term on the right hand side, which does not have such dependence,
the S$^5$ dependence has to cancel as well.

\subsubsection{Solving for \texorpdfstring{S$^2$}{S2} dependences}
From the fact that the S$^5$ dependence in the first term on the right hand side of (\ref{eqn:lin-kappa-2}) has to cancel,
we now see that either $f$ and $\rho$ have to have the same dependence on $r$ up to overall constants,
or $\star d\omega$ has to be constant.
The first option is the one leading to non-trivial results and we thus write $\rho=f\psi$, where $f$ is the radial profile that appeared already in (\ref{eq:A-lin-split}).
Note that the relative normalization of $\omega$ and $\psi$ now matters -- they can not both be chosen normalized w.o.l.g.\ at the same time.
Using also $\Gamma_{11}\epsilon^{(0)}=\epsilon^{(0)}$ to convert $\GammaAdS$ to $\GammaS$,
we then find
\begin{align}\label{eqn:lin-kappa-3}
\begin{split}
 \Big[\theta^\prime\Gamma_{\underline{\rho}}\Gamma_{\underline{r}}+i\theta\Gamma_{\underline{\rho}}\GammaAdS\Big]\epsilon^{(0)}
 &=
 f\Big[
 i\big(\psi-\star d\omega\big)\GammaS-(\partial_1\psi)\Gamma_{\underline{\beta_1}}-\csc\beta_1(\partial_2\psi)\Gamma_{\underline{\beta_2}}
 \Big]\Gamma_{\underline{\theta}}\epsilon^{(0)}
 \\&
 \hphantom{=}\ +f^\prime\Big[
 i\omega_2\csc\beta_1\Gamma_{\underline{\beta_1}}-i \omega_1\Gamma_{\underline{\beta_2}}
 +\psi\GammaS
 \Big]\Gamma_{\underline{\theta}}\Gamma_{\underline{r}}\GammaAdS\epsilon^{(0)}~.
\end{split}
\end{align}
We start with the second term on the right hand side, in which all S$^5$ dependence has to drop out after multiplying by $\RS^{-1}$,
as argued above.
Note that this term is algebraic in $\omega$ and $\psi$.
To make the structure more apparent, we rewrite (\ref{eqn:lin-kappa-3}) as
\begin{align}\label{eqn:lin-kappa-4}
 \Big[\theta^\prime\Gamma_{\underline{\rho}}\Gamma_{\underline{r}}+i\theta\Gamma_{\underline{\rho}}\GammaAdS\Big]\epsilon^{(0)}
 &=
 f Y \epsilon^{(0)} +f^\prime Z\, \Gamma_{\underline{r}}\GammaAdS\epsilon^{(0)}~,
\end{align}
where $Y$ and $Z$ are independent of $r$, and are given by
\begin{subequations}
\begin{align}
 Y&=\Big[i\big(\psi-\star d\omega\big)\GammaS-(\partial_1\psi)\Gamma_{\underline{\beta_1}}-\csc\beta_1(\partial_2\psi)\Gamma_{\underline{\beta_2}}\Big]\Gamma_{\underline{\theta}}~,
 \\
 Z&=\Big[i\omega_2\csc\beta_1\Gamma_{\underline{\beta_1}}-i \omega_1\Gamma_{\underline{\beta_2}}
 +\psi\GammaS\Big]\Gamma_{\underline{\theta}}~.
\end{align}
\end{subequations}
So the task is to find $\psi$, $\omega$ such that $\RS^{-1}Z\RS$ is independent of the S$^5$ coordinates.
We can see immediately that $\RS^{-1}Y\RS$ and $\RS^{-1}Z\RS$ will involve the same Clifford algebra structures.
For the explicit evaluation we will fix the position at which the D5-branes are located on the $\tilde S^2$ in (\ref{eqn:metric-AdS5}) to $\alpha_1=0$.
This can be done without loss of generality, since we can arbitrarily choose which point corresponds to the north pole in the coordinates $\alpha_1,\alpha_2$.
At $\theta=0$ we then find
\begin{subequations}
\begin{align}
 \RS^{-1}\Gamma_{\underline{\beta_1}}\Gamma_{\underline{\theta}}\RS&=
 \cos\beta_1\big(\cos\beta_2\Gamma_{\underline{\beta_1}}+\sin\beta_2\Gamma_{\underline{\beta_2}}\big)\Gamma_{\underline{\theta}}
 -\sin\beta_1\GammaS\Gamma_{\underline{\theta}}~,
 \\
 \RS^{-1}\Gamma_{\underline{\beta_2}}\Gamma_{\underline{\theta}}\RS&=
 \big(\cos\beta_2\Gamma_{\underline{\beta_2}}-\sin\beta_2\Gamma_{\underline{\beta_1}}\big)\Gamma_{\underline{\theta}}~,
 \\
 \RS^{-1}\GammaS\Gamma_{\underline{\theta}}\RS&=
 \cos\beta_1\GammaS\Gamma_{\underline{\theta}}
 -i\sin\beta_1\big(
 \cos\beta_2\Gamma_{\underline{\beta_1}}+\sin\beta_2\Gamma_{\underline{\beta_2}}
 \big)\Gamma_{\underline{\theta}}~.
\end{align}
\end{subequations}
The expressions involve three independent Clifford algebra structures, and to eliminate the S$^5$ dependence we have to solve
\begin{align}\label{eqn:omega-psi-eq}
 \RS^{-1}Z\RS&=\big(i c_1\Gamma_{\underline{\beta_1}}+i c_2\Gamma_{\underline{\beta_2}}+c_3\GammaS\big)\Gamma_{\underline{\theta}}~,
\end{align}
with generically complex constants $c_1$, $c_2$, $c_3$.
This is a system of three linear equations for $\omega_i$ and $\psi$, and can be solved straightforwardly.
Demanding $\RS^{-1}Y\RS$ to be independent of the position on S$^5$ as well adds another three equations, overconstraining the system.
The system we get from (\ref{eqn:omega-psi-eq}) is
\begin{align}\label{eqn:S2-dependence-eq}
 \begin{pmatrix}
 -\sin\beta_1\cos\beta_2 & \sin\beta_2 & \cot\beta_1 \cos\beta_2\\
 -\sin\beta_1\sin\beta_2 & -\cos\beta_2 & \cot\beta_1\sin\beta_2\\
 \cos\beta_1 & 0 & 1
 \end{pmatrix}
 \begin{pmatrix}\psi\\ \omega_1\\ \omega_2\end{pmatrix}
 &=
 \begin{pmatrix}c_1\\ c_2\\c_3\end{pmatrix}~.
\end{align}
For all $c_i$ real the solutions are real, and we shall assume that henceforth.
The general solution reads
\begin{subequations}\label{eqn:S2-dependence-sol}
\begin{align}
 \psi&=c_3 \cos\beta_1-\sin\beta_1(c_1 \cos\beta_2+c_2\sin\beta_2)~,\\
 \omega_1&=c_1\sin\beta_2-c_2\cos\beta_2~,\\
 \omega_2&=\sin\beta_1(\cos\beta_1(c_1 \cos\beta_2+c_2\sin\beta_2)+c_3\sin\beta_1)~.
\end{align}
\end{subequations}
It satisfies $\bigtriangleup_{\mathrm{S}^2}\psi=-2\psi$, so $\psi$ is an $\ell=1$ mode.
Furthermore, $\omega_i=\sqrt{g_{\mathrm{S}^2}}\varepsilon_{ij}g_{\mathrm{S}^2}^{jk}\partial_k\psi$ with $\varepsilon_{12}=1$,
so we also have $\psi=\frac{1}{2}\star d\omega$.
For $c_1=c_2=0$ there is no $\beta_2$ dependence and translations in $\beta_2$ are the preserved U(1).
Finally, we have to verify that this solution also makes $\RS^{-1}Y\RS$ independent of the S$^5$ coordinates.
Indeed, using (\ref{eqn:S2-dependence-sol}) results in 
\begin{align}
 \RS^{-1}Y\RS&=-i\RS^{-1}Z\RS~.
\end{align}
So the S$^5$ dependence drops out in the first term on the right hand side of (\ref{eqn:lin-kappa-4}) as well, as desired.

\subsubsection{Solving for the radial profiles}
With the results of the previous section we have eliminated all non-trivial dependence on the $\rm S^5$ and can focus on the $\rm AdS_5$ part.
This will yield the explicit form of the necessary projector on the Killing spinor, singling out the  preserved supersymmetries, and the linearized BPS equations for infinitesimally massive embeddings.
Eq.~(\ref{eqn:lin-kappa-4}), after multiplying both sides by $\RS^{-1}$ and using (\ref{eqn:S2-dependence-sol}), becomes
\begin{align}\label{eqn:lin-kappa-5}
 \Big[\theta^\prime\Gamma_{\underline{\rho}}\Gamma_{\underline{r}}+i\theta\Gamma_{\underline{\rho}}\GammaAdS\Big]\RAdS^{(0)}\epsilon_0
 &=\Big[f\, \mathds{1}+if^\prime \Gamma_{\underline{r}}\GammaAdS\Big]\RAdS^{(0)}\Gamma_p\epsilon_0~,
\end{align}
where we have defined
\begin{align}\label{eq:Gamma-p}
 \Gamma_p&=\big(c_1\Gamma_{\underline{\beta_1}}+c_2\Gamma_{\underline{\beta_2}}-i c_3\GammaS\big)\Gamma_{\underline{\theta}}~.
\end{align}
Note that $\Gamma_p^2=-|c|^2\mathds{1}$ with $|c|^2=c_1^2+c_2^2+c_3^2$, and $\Gamma_p$ commutes with AdS$_5$ $\Gamma$-matrices.
So it can be used straightforwardly for constructing projectors.

To solve (\ref{eqn:lin-kappa-5}) systematically we turn to the dependence on the AdS$_3$ directions $(x,t,y)$.
We introduce an operator $\mathcal R_A$ and for later convenience also $\mathcal R_S$, defined by
\begin{align}\label{eqn:Rop}
 \mathcal R_A[\Gamma]&=\RAdS^{-1}\Gamma\RAdS~,
 &
 \mathcal R_S[\Gamma]&=\RS^{-1}\Gamma\RS~.
\end{align}
At $\rho=0$, we then note the useful identity
\begin{align}
 \mathcal R_A^{(0)}[\Gamma_{\underline{\rho}}\Gamma_{\underline{r}}]&=
 i\tanh r \,\mathcal R_A^{(0)}[\Gamma_{\underline{\rho}}\GammaAdS]+\sech r\,\Gamma_{\underline{\rho}}\Gamma_{\underline{r}}~.
\end{align}
Multiplying (\ref{eqn:lin-kappa-5}) by $(\RAdS^{(0)})^{-1}$, we then find that the $\kappa$-symmetry condition becomes
\begin{align}\label{eqn:lin-kappa-6}
 i\big(\theta^\prime\tanh r+\theta\big)\mathcal R_A^{(0)}[\Gamma_{\underline{\rho}}\GammaAdS]\epsilon_0
 +\theta^\prime\sech r\,\Gamma_{\underline{\rho}}\Gamma_{\underline{r}}
 &=
 f\Gamma_p\epsilon_0+i f^\prime\mathcal R_A^{(0)}[\Gamma_{\underline{r}}\GammaAdS]\Gamma_p\epsilon_0~.
\end{align}
The parts without $\mathcal R_A$ already suggest a form of the projector to impose, but since they may be modified by contributions from the terms involving $\mathcal R_A$,
we need to analyze the $(x,t,y)$ dependent terms to derive the projector.
The explicit identities we will use for that are
\begin{subequations}
\begin{align}\label{eqn:calR-identities}
 \mathcal R_A^{(0)}[\Gamma_{\underline{r}}\GammaAdS]&=
 ie^x\big(y\Gamma_{\underline{y}}+t\Gamma_{\underline{t}}\big)\Gamma_{\underline{r}}
 -i\big(e^x(y^2-t^2-1)+e^{-x}\big)\Gamma_{\underline{x}}P_{x+}\Gamma_{\underline{r}}+e^{x}\Gamma_{\underline{r}}\GammaAdS~,
 \\
 \mathcal R_A^{(0)}[\Gamma_{\underline{\rho}}\GammaAdS]\Gamma_{\underline{\rho}}\Gamma_{\underline{r}}&=
 \cosh r\,\mathcal R_A^{(0)}[\Gamma_{\underline{r}}\GammaAdS]+i\sinh r\,\mathds{1}~.
\end{align}
\end{subequations}
The first one shows that the AdS$_3$ dependence can not be canceled within $\mathcal R_A^{(0)}[\Gamma_{\underline{\rho}}\GammaAdS]$
and $R_A^{(0)}[\Gamma_{\underline{r}}\GammaAdS]$ separately, since there are too many independent Clifford algebra structures with non-trivial dependence.
The second identity shows precisely which projector has to be imposed on $\epsilon_0$ to have a chance to cancel the non-trivial AdS$_3$ dependence between the two terms.
Using it in (\ref{eqn:lin-kappa-6}) gives
\begin{align}
\label{eqn:kappa-lin-proj}
 i\mathcal R_A^{(0)}[\Gamma_{\underline{\rho}}\GammaAdS]\big(\theta^\prime\tanh r+\theta-f^\prime\sech r\,\Gamma_{\underline{\rho}}\Gamma_{\underline{r}}\Gamma_p\big)\epsilon_0
 &=
 \big(f+f^\prime\tanh r\big)\Gamma_p\epsilon_0-\theta^\prime\sech r\,\Gamma_{\underline{\rho}}\Gamma_{\underline{r}}\epsilon_0~.
\end{align}
From this equation we can finally read off the projector, from the requirement to make the left hand side cancel in a non-trivial way. 
This yields
\begin{align}\label{eqn:massive-projector}
 \frac{1}{|c|}\Gamma_{\underline{\rho}}\Gamma_{\underline{r}}\Gamma_p\epsilon_0&=\lambda \epsilon_0~,
 &
 \lambda&=\pm 1~,
\end{align}
with $\Gamma_p$ defined in (\ref{eq:Gamma-p}).
With this projector the equation for $\kappa$-symmetry in (\ref{eqn:kappa-lin-proj}) becomes
\begin{align}
 i\mathcal R_A^{(0)}[\Gamma_{\underline{\rho}}\GammaAdS]\big(\theta^\prime\tanh r+\theta-\lambda |c| f^\prime\sech r\big)\epsilon_0
 &=
 -\big(\theta^\prime\sech r+\lambda |c|(f+f^\prime\tanh r)\big)\Gamma_{\underline{\rho}}\Gamma_{\underline{r}}\epsilon_0~.
\end{align}
The terms in brackets on each side now have to vanish separately and we obtain two 1$^\mathrm{st}$-order equations for the slipping and bending modes
\begin{align}
 \theta^\prime\tanh r+\theta-\lambda |c| f^\prime\sech r&=0~,
 &
 \theta^\prime\sech r+\lambda |c|(f+f^\prime\tanh r)&=0~.
\end{align}
They have non-trivial solutions and we have 
verified that these two equations indeed 
imply the linearized equations of motion derived from the action in (\ref{eqn:D5-action}).

\subsection{Non-linear \texorpdfstring{$\kappa$}{kappa}-symmetry}\label{sec:non-linear-kappa}
In this section we derive the form of the fully non-linear $\kappa$-symmetry equations for embeddings preserving the eight supersymmetries 
that are preserved by massive defect fields.
We have the general form of the $\kappa$-symmetry condition in (\ref{eqn:kappa-condition}) and the two projection conditions on the background 
Killing spinors derived in the previous section and given in (\ref{eqn:massless-projector}) and (\ref{eqn:massive-projector}).
Note that the projection conditions in (\ref{eqn:massless-projector}) and (\ref{eqn:massive-projector}) restrict the constant spinor $\epsilon_0$ 
that parametrizes the $\rm AdS_5\times S^5$ Killing spinors via (\ref{eqn:Killing-R}). 
The conditions therefore carry over directly from the infinitesimally massive case to the general case considered now, where the $R$-matrices in (\ref{eqn:Killing-R})
evaluated at the location of the D5-branes may take a different form.

We start out from (\ref{eqn:kappa-condition}) and handle the first term in that equation first.
With the definition of the charge conjugated $R$-matrices in (\ref{eqn:Rtilde-identities}) we find
\begin{align}
 C\epsilon^\star&=-e^{-i\rho\Gamma_{\underline{\rho}}\GammaAdS}e^{-i\theta\Gamma_{\underline{\theta}}\GammaS}\Gamma_{\underline{\rho}}\GammaUbeta
 \RAdS\RS\Gamma_{\underline{\rho}}\GammaUbeta\,C(\epsilon_0)^\star~.
\end{align}
We can now use the massless projection condition (\ref{eqn:massless-projector}) to eliminate the charge-conjugated constant spinor, and find
\begin{align}
 C\epsilon^\star&=ie^{-i\rho\Gamma_{\underline{\rho}}\GammaAdS}e^{-i\theta\Gamma_{\underline{\theta}}\GammaS}\Gamma_{\underline{\rho}}\GammaUbeta\GammaAdS\epsilon~.
\end{align}
The $\kappa$-symmetry condition (\ref{eqn:kappa-condition}) then involves no more charge conjugation and becomes
\begin{align}\label{eqn:kappa-nonlinear1}
 -\hat\Gamma e^{-i\rho\Gamma_{\underline{\rho}}\GammaAdS}e^{-i\theta\Gamma_{\underline{\theta}}\GammaS}\Gamma_{\underline{\rho}}\GammaUbeta\GammaAdS\epsilon
 +\frac{i}{2}\gamma^{ij}F_{ij}\hat\Gamma\epsilon&=h\epsilon~.
\end{align}
For the gauge field term it is convenient to use the field strength with indices raised by the induced metric on the D5-brane worldvolume, and write it as
\begin{align}\label{eqn:gamma-F}
 \frac{1}{2}\gamma^{ij}F_{ij}=\gamma_{r\beta_i}F^{r\beta_i}+\gamma_{\beta_1\beta_2}F^{\beta_1\beta_2}~.
\end{align}
The remaining task is to use the massive projector in (\ref{eqn:massive-projector}) to derive an explicit form for the conditions implied by the constraint (\ref{eqn:kappa-nonlinear1}).
This is a cumbersome but straightforward task, and we give the details in app.~\ref{app:non-linear-kappa}.
The result is that, after multiplying eq.~(\ref{eqn:kappa-nonlinear1}) by $\RAdS^{-1}\RS^{-1}$, the condition can be written as
\begin{align}
\mathcal Q_{\mathcal K}\,\mathcal K\Gamma_{\underline{\rho}}\epsilon_0 
+\mathcal Q_{\mathds{1}}\epsilon_0&=h\epsilon_0~,
\end{align}
where $\mathcal K$ is a matrix with explicit dependence on the AdS$_3$ coordinates,
and the explicit form of $\mathcal Q_{\mathcal K}$ and $\mathcal Q_{\mathds{1}}$ can be found in app.~\ref{app:non-linear-kappa}.
The crucial point is that the available projectors and chirality constraints have been implemented completely, 
such that $\mathcal Q_{\mathcal K}$ and $\mathcal Q_{\mathds{1}}$ necessarily have to vanish separately for the $\kappa$-symmetry condition to be satisfied.
We thus have to solve
\begin{align}\label{eqn:kappa-non-linear-B}
 \mathcal Q_{\mathcal K}&=0~,
 &
 \mathcal Q_{\mathds{1}}- h\mathds{1}&=0~.
\end{align}
We will discuss the explicit form of the implied equations in the next section.

\subsection{The BPS equations}
In this section we spell out explicitly the equations resulting from the two matrix equations in (\ref{eqn:kappa-non-linear-B}).
For the massive projector in (\ref{eqn:massive-projector}) we left the choice of $c_1$, $c_2$ and $c_3$ arbitrary up to this point.
But we shall from now on set 
\begin{align}\label{eq:c-choice}
 c_1&=c_2=0~, &c_3=1~.
\end{align}
There is no explicit dependence on $\beta_2$ in the metric and with this choice of projector the Clifford algebra manipulations do not introduce dependence either,
such that the U(1) isometry remaining from the $\rm S^2$ that the D5-branes wrap in $\rm S^5$ is realized as translations in $\beta_2$.
The equations can then be solved with only trivial dependence on $\beta_2$.
Moreover, the first equation in (\ref{eqn:kappa-non-linear-B}) in particular implies 
$\tr \big(\tilde {\mathcal Q}_{\mathcal{K}}\Gamma_{\underline{\theta\beta_1\beta_2}}\big)=0$,
and with the specific choice of projector in (\ref{eq:c-choice}) this condition implies 
\begin{align}
F_{r\beta_1}&=0~.
\end{align}
That means $A_{\beta_1}$ is independent of $r$ and depends on $\beta_1$ only, and can be set to zero by a residual gauge transformation.

To conveniently write the remaining equations resulting from (\ref{eqn:kappa-non-linear-B}) we change coordinates to
\begin{align}\label{eqn:coord-transf}
 r&=2\tanh^{-1}\tan\frac{z}{2}~,
 &
 \beta_1&=\cos^{-1}x~,
\end{align}
such that $z\in(-\frac{\pi}{2},\frac{\pi}{2})$ to completely cover $\rm AdS_5$, and $x\in[0,1]$.
As shorthands we also introduce
\begin{align}
F_{z}&:=F_{z\beta_2}=\frac{\partial r}{\partial z}F_{r\beta_2}~, 
&
F_x&:=F_{x\beta_2}=\frac{\partial \beta_1}{\partial x}F_{\beta_1\beta_2}~.
\end{align}
Moreover, with $\mathfrak{a},\mathfrak{b}=z,x$ and $\epsilon^{zx}=1$,
we define
\begin{align}
 B_1&=\varepsilon^{\mathfrak{ab}}(\partial_{\mathfrak{a}}\theta) F_{\mathfrak{b}}~,
 &
 B_2&=\varepsilon^{\mathfrak{ab}}(\partial_{\mathfrak{a}}\rho) F_{\mathfrak{b}}~,
 &
 B_3&=\varepsilon^{\mathfrak{ab}}(\partial_{\mathfrak{a}}\rho)\partial_{\mathfrak{b}}\theta~,
\end{align}
and
\begin{subequations}
\begin{align}
 A_1&=\partial_x(\sinh\rho\cos\theta)~,&
 A_2&=\partial_x(\sinh\rho\sin\theta)~,\\
 A_3&=\sech\rho (\lambda x \partial_z \theta+\tan z\partial_z\rho)-\sec^2 z\sinh\rho~,&
 A_4&=\partial_x(\cosh\rho\cos\theta)~,\\
 A_5&=\partial_x(\cosh\rho\sin\theta)~.
\end{align}
\end{subequations}
From $\mathcal Q_{\mathcal K}=0$ we find four equations, which, with the shorthands defined above, take the form
\begin{subequations}\label{eq:QK2}
\begin{align}
 \cos^2\theta\big(\sin\theta A_3+\lambda\tanh\rho F_z\big)+\tan z B_1-\lambda x B_2&=0~,
 \label{eq:QK2-1}
 \\
 \lambda\partial_z\sin\theta+\sin\theta\tan z B_3-\lambda\sinh\rho B_1
 +\cosh\rho\Big[\frac{F_z\tan z}{1-x^2}-\lambda\tan\theta B_2-\sec^2 zA_1\Big]&=0~,
 \\
 \lambda\sinh\rho\Big[\frac{x F_z}{1-x^2}-\tan\theta B_1\Big]-\tan z \cos\theta B_3+\cosh\rho\big(\lambda B_2+\sec^2z A_2\big)&=0~,
 \label{eq:QK2-3}
 \\
 \cos^2\theta\big(\cos\theta A_3-\lambda\tan\theta\tanh\rho F_z\big)-\tanh\rho\big(\lambda x B_1+\tan z B_2\big)-\sec^2z F_x&=0~.
 \label{eq:QK2-4}
\end{align}
\end{subequations}
We similarly find four equations resulting from $\mathcal Q_{\mathds{1}}-h\mathds{1}=0$, and those read
\begin{subequations}\label{eq:Qid2}
\begin{align}
\begin{split}
 -\frac{h}{\sqrt{1-x^2}}+\cos\theta\cosh^2\rho\Big[\cos^2\theta-\lambda\tanh\rho\tan z\Big((1-x^2)\partial_x\theta+\frac{x}{2}\sin2\theta\Big)\Big]
 \\ 
 +\frac{\lambda}{2}\sin 2\theta\big(x\cos\theta\partial_z\rho+\cosh\rho F_z\big)
 +\lambda x\cosh\rho B_1
 +\lambda(1-x^2)\cos\theta B_3
 &=0~,
\end{split}
\label{eq:Qid2-1}
 \\
\begin{split}
 \cosh\rho\Big[\frac{x F_z}{1-x^2}-\tan\theta B_1+\tan z\big(F_x+x A_1+\sinh\rho\cos\theta\big)+\lambda A_5\Big]
 \\
 -\cos\theta\partial_z\rho+\sinh\rho B_2+x\sin\theta B_3&=0~,
\end{split}
\label{eq:Qid2-2}
\\
\lambda\tan z\big(\tan\theta F_x+x A_2\big)
-A_4
+\lambda\big(\tan\theta\tanh\rho B_2+B_1-x\sech\rho\cos\theta B_3\big)&=0~,
\label{eq:Qid2-3}
\\
\begin{split}
\lambda x \cos^3\theta\big(\sech\rho\partial_z\rho-\tan z\sinh\rho\big)-\lambda x \tanh\rho B_2
\\
+\cos^2\theta\big(\lambda F_z-\sin\theta\cosh\rho\big)
+\lambda\tan z\big((1-x^2)\cos\theta\partial_x\sinh\rho-x F_x)&=0~.
\end{split}
\label{eq:Qid2-4}
\end{align}
\end{subequations}
We therefore have eight equations for two functions $\rho$, $\theta$ and the two field strength components $F_{r\beta_2}$, $F_{\beta_1\beta_2}$,
which in addition have to satisfy the Bianchi identity for $F$.
The equations are clearly non-linear (in the functions and in derivative terms), and we have a square root implicit in the definition of $h$ in eq.~(\ref{eq:Qid2-1}).
However, except for (\ref{eq:Qid2-1}) the equations are only quadratic in derivative terms, and 
since we have more equations than functions we may attempt to derive quasilinear equations by taking combinations of these equations.
This will be done in the next section.

\subsection{Solving for the gauge field}\label{sec:solving-A}
In this section we will use part of the BPS equations (\ref{eq:QK2}) and (\ref{eq:Qid2}) to eliminate the terms quadratic in derivatives, 
to find equations which are linear in derivative terms.
These can then be used to solve for the remaining component of the gauge field, $A_{\beta_2}$.
Eqs.~(\ref{eq:QK2-1})-(\ref{eq:QK2-3}) can be regarded as a set of linear equations for $(B_1,B_2,B_3)$ and we can solve them for the terms quadratic in derivatives.
This yields
\begin{subequations}\label{eq:solB}
\begin{align}
 \sinh\rho B_1&=\cos\theta\Big[\lambda\sec^2z\cosh^2\rho\partial_x\rho+\cos^2\theta\partial_z\theta
 +\frac{F_z}{1-x^2}C_+\Big]~,
 \label{eq:solB1}
 \\
 x\sec^2\theta B_2&=\coth\rho\tan z\sec^2 z\big(A_1+A_2\tan\theta\big)+\lambda A_3\sin\theta+\lambda\csch\rho\tan z \partial_z\sin\theta
 \nonumber\\
 &\hphantom{=}+F_z\tanh\rho+\frac{\lambda F_z}{1-x^2}\sec\theta\csch\rho\tan z C_+~,
 \label{eq:solB2}
 \\
 x\sec\theta B_3&=
 \cosh\rho\big(x\csc z\sec z(A_2-\tan\theta A_1)+ A_3\sin\theta\cot z+\lambda\coth\rho\sec^2z(A_1+A_2\tan\theta)\big)
 \nonumber\\
 &\hphantom{=}-\lambda\csch\rho\cot z C_-\partial_z\theta+\frac{\lambda F_z}{1-x^2}(\sinh\rho\cot z+\cosh\rho\coth\rho\tan z)~,
 \label{eq:solB3}
\end{align}
\end{subequations}
where
\begin{align}
 C_\pm&=x\sinh\rho\sin\theta\pm\lambda\tan z\cos\theta\cosh\rho~.
\end{align}
We will not use (\ref{eq:Qid2-1}) to avoid introducing square roots,
and using the solution for $(B_1,B_2,B_3)$ thus leaves us with four equations which are linear in derivative terms, namely (\ref{eq:QK2-4})
and (\ref{eq:Qid2-2})-(\ref{eq:Qid2-4}).
A linear combination which is particularly helpful can be isolated as follows.
We solve (\ref{eq:QK2-4}) and (\ref{eq:Qid2-2}), with $B_i$ replaced according to (\ref{eq:solB}), for $\partial_x\rho$ and $\partial_x\theta$, 
and use the result in (\ref{eq:Qid2-3}) and (\ref{eq:Qid2-4}). They then both become
\begin{align}\label{eq:Fz}
 \partial_z\big(x\sinh\rho\cos\theta-\lambda \tan z \cosh\rho\sin\theta\big)+F_z&=0~.
\end{align}
This equation can be integrated straightforwardly for the gauge potential, which has to be given by the expression in the round brackets up to a function of $x$ only.
We can now use this result for $F_z$ in eq.~(\ref{eq:QK2-4}), solve for $\partial_z\theta$ and use the result in (\ref{eq:Qid2-2}).
This yields
\begin{align}\label{eq:Fx}
 \partial_x\big(x\sinh\rho\cos\theta-\lambda \tan z \cosh\rho\sin\theta\big)+F_x&=0~.
\end{align}
Together with (\ref{eq:Fz}) this fixes the solution for the gauge field to 
\begin{align}\label{eq:solA}
 A_{\beta_2}&=\lambda\tan z \cosh\rho\sin\theta-x\sinh\rho\cos\theta+\mathcal A_0~,
\end{align}
with an arbitrary constant $\mathcal A_0$. The Bianchi identity for $F$ is then automatically satisfied.
We will derive one more useful quasilinear equation before collecting and discussing the complete set of remaining equations in the next subsection.
To this end, we use the solution for the gauge field, solve eq.~(\ref{eq:QK2-4}) for $\partial_z\theta$ and use the result in (\ref{eq:Qid2-3}).
The resulting equation is
\begin{align}\label{eq:rho-theta-lin}
 (1-x^2)\cosh\rho\big(\lambda x \partial_x\theta+\tan z \partial_x\rho-\lambda\sin\theta\cos\theta\big)+\cos\theta F_z&=0~.
\end{align}

\subsection{The remaining equations}\label{sec:remainingBPS}
We will now collect and discuss the complete set of remaining equations after solving for the gauge field.
Upon using the solution for the gauge field (\ref{eq:solA}) and the quasilinear equation (\ref{eq:rho-theta-lin}), 
the $\mathcal Q_\mathcal{K}=0$ equations (\ref{eq:QK2-1})-(\ref{eq:QK2-4}) and
the last three of the $\mathcal Q_{\mathds{1}}-h\mathds{1}=0$ equations (\ref{eq:Qid2-2})-(\ref{eq:Qid2-4}) all become equivalent.
This may be verified by solving (\ref{eq:rho-theta-lin}) for $\partial_x\theta$ and replacing it everywhere, after using (\ref{eq:solA}).
This leaves only the first of the $\mathcal Q_{\mathds{1}}-h\mathds{1}=0$ equations, (\ref{eq:Qid2-1}), and (\ref{eq:rho-theta-lin}) in addition.
We are thus left with three equations for two functions $\rho$, $\theta$, which we repeat for convenience.
Of the now equivalent eqs.~(\ref{eq:QK2-1})-(\ref{eq:QK2-4}) and (\ref{eq:Qid2-2})-(\ref{eq:Qid2-4}), we pick (\ref{eq:QK2-4}), 
leaving us with
\begin{align}
 (1-x^2)\cosh\rho\big(\lambda x \partial_x\theta+\tan z \partial_x\rho-\lambda\sin\theta\cos\theta\big)+\cos\theta F_z&=0~,
 \label{eq:rho-theta-lin-rep}
 \\
  \cos^2\theta\big(\cos\theta  A_3-\lambda\tan\theta\tanh\rho F_z\big)-\tanh\rho\big(\lambda x B_1+\tan z B_2\big)-\sec^2z F_x&=0~,
 \label{eq:QK2-4-rep}
 \\
 \begin{split}
 -\frac{h}{\sqrt{1-x^2}}+\cos\theta\cosh^2\rho\Big[\cos^2\theta-\lambda\tanh\rho\tan z\Big((1-x^2)\partial_x\theta+\frac{x}{2}\sin2\theta\Big)\Big]
 \\ 
 +\frac{\lambda}{2}\sin 2\theta\big(x\cos\theta\partial_z\rho+\cosh\rho F_z\big)
 +\lambda x\cosh\rho B_1
 +\lambda(1-x^2)\cos\theta B_3
 &=0~.
\end{split}
\label{eq:Qid2-1-rep}
\end{align}
These three equations imply all the others and the Bianchi identity.
We will now show that the third equation, (\ref{eq:Qid2-1-rep}), is actually also implied already by (\ref{eq:rho-theta-lin-rep}) and (\ref{eq:QK2-4-rep}).
Using the shorthands $B_i$ and the definition of $h$, we can write $h^2$ as
\begin{align}\label{eqn:h2}
\begin{split}
 \frac{h^2}{1-x^2}&=
 \cos ^4\theta \cosh ^2\rho+
 \cos ^2z \left[B_1^2+B_2^2+(1-x^2) \cos ^2\theta B_3^2 +\big((\partial_z\theta)^2+(\partial_z\rho)^2\big) \cos ^4\theta\right]
   \\&\hphantom{=}
   +\left(1-x^2\right) \left[(\partial_x\theta)^2+(\partial_x\rho)^2\right] \cos ^2\theta \cosh^2\rho+F_x^2 \cosh ^2\rho
   +\frac{F_z^2\cos ^2\theta \cos ^2z}{1-x^2}~.
\end{split}
\end{align}
We now isolate the term involving $h$ in (\ref{eq:Qid2-1-rep}) and square the equation afterwards,
to eliminate the square root.
This results in
\begin{align}\label{eq:Qid2-1-rep-squared}
\begin{split}
 \frac{h^2}{1-x^2}&=\bigg[
 \cos\theta\cosh^2\rho\Big[\cos^2\theta-\lambda\tanh\rho\tan z\Big((1-x^2)\partial_x\theta+\frac{x}{2}\sin2\theta\Big)\Big]
 \\ &\hphantom{=\Big[}
 +\frac{\lambda}{2}\sin 2\theta\big(x\cos\theta\partial_z\rho+\cosh\rho F_z\big)
 +\lambda x\cosh\rho B_1
 +\lambda(1-x^2)\cos\theta B_3
 \bigg]^2~.
\end{split}
\end{align}
With the expression for $h^2$ in (\ref{eqn:h2}) we see that this equation is quadratic in the $B_i$, and thus quartic in derivative terms.
However, using (\ref{eq:solB}) reduces it to an equation which is only quadratic in derivative terms.
To further process it, we  solve (\ref{eq:rho-theta-lin-rep}) for $\partial_x\theta$ and eliminate it in 
(\ref{eq:QK2-4-rep}) and in (\ref{eq:Qid2-1-rep-squared}).
Eq.~(\ref{eq:QK2-4-rep}) is quadratic in derivative terms, and after eliminating $\partial_x\theta$ via (\ref{eq:rho-theta-lin-rep}) it contains a $(\partial_z\rho)(\partial_z\theta)$ term. 
We solve (\ref{eq:QK2-4-rep}) for $(\partial_z\rho)(\partial_z\theta)$ and use the result to eliminate that
particular quadratic derivative term in (\ref{eq:Qid2-1-rep-squared}) with the $B_i$ replaced according to (\ref{eq:solB}).
After this step (\ref{eq:Qid2-1-rep-squared}) collapses to zero, showing that the equation is implied by (\ref{eq:rho-theta-lin-rep}) and (\ref{eq:QK2-4-rep}).

The remaining equations are therefore only (\ref{eq:rho-theta-lin-rep}) and (\ref{eq:QK2-4-rep}).
These two equations for slipping and bending mode, together with the solution for the gauge field
(\ref{eq:solA}), are equivalent to the entire set of 
equations in (\ref{eq:QK2}) and (\ref{eq:Qid2}).
We rewrite them as follows.
With
\begin{align}
 G_{\mathfrak{a}}&=\lambda\sinh\rho\sin^2\theta\,\partial_{\mathfrak{a}}(x\cot\theta)-\partial_{\mathfrak{a}}(\tan z \cosh\rho)~,
\end{align}
they become
\begin{subequations}\label{eq:BPS}
\begin{align}
 -(1-x^2)\cosh\rho\, G_x+\sinh\rho\cos\theta F_z&=0~,
 \label{eq:rho-theta-lin-rep-2}
 \\
 \sinh\rho\,(G_z F_x-G_x F_z)-\cos^3\!\theta\, G_z -\sec^2\!z\, \cosh^3\!\rho\,\cos^3\!\theta&=0~.
 \label{eq:QK2-4-rep-2}
\end{align}
\end{subequations}
We have not been able to find a closed-form solution to these equations.
But to validate that the equations are correct, we solved them to cubic order in a perturbative expansion around the
straightforward solution where $\rho$ and $\theta$ vanish identically, and verified that these perturbative solutions solve
the equations of motion resulting from the DBI action in (\ref{eqn:D5-action}).
Since the cubic order in the perturbative expansion is sensitive to all terms in the equations, this provides a strong consistency check on the full non-linear equations.
We will study physical applications of the perturbative solutions in more detail in the next sections.

\section{Perturbative solutions for \texorpdfstring{$\rm AdS_3$}{AdS3} defects}\label{sec:general-linearized}
In this section we study perturbative solutions to the BPS equations derived in the previous section. 
We start with the general linearized solution, which describes infinitesimal fluctuations around the massless embedding 
where $\rho$, $\theta$ and $A_{\beta_2}$ vanish identically.
We will find a large space of supersymmetric 'vacuum states', where the expectation values of all the supersymmetric operators sourced by combinations of slipping mode, bending mode and gauge field can be varied continuously and independently on one of the AdS$_3$ geometries obtained as $r\rightarrow \pm\infty$.
Once the expectation values are fixed on one of the AdS$_3$ patches, they are then fixed on the remaining one as well.
We will close with a qualitative discussion of higher orders in the perturbative expansion and of implications of that general structure for non-linear solutions.

To make the expansion explicit we introduce a small parameter $\kappa$ and expand around the massless embedding in the form
\begin{align}\label{eq:pert-exp}
 \rho&=\sum_n\kappa^n\rho^{(n)}~,
 &
 \theta&=\sum_n\kappa^n\theta^{(n)}~.
\end{align}
The solution for the gauge field (\ref{eq:solA}) can be expanded in a similar way and at linear order becomes
\begin{align}\label{eq:solA3}
 A_{\beta_2}^{(1)}&=\lambda\tan z\,\theta^{(1)} - x \rho^{(1)}+\mathcal A_0^{(1)}~.
\end{align}
With only $A_{\beta_2}$ non-vanishing, we automatically have $d^\dagger_{\mathrm{S}^2}A=0$,
or $\nabla^{i}A_i=0$ with $i$ running over the S$^2$ indices corresponding to $(\beta_1,\beta_2)$ only.
The linearized versions of (\ref{eq:BPS}) read
\begin{subequations}\label{eqn:lin-eq}
\begin{align}
 \partial_z A_{\beta_2}^{(1)}+(1-x^2)\partial_x\big(\lambda x \theta^{(1)}+\tan z\,\rho^{(1)})-2\lambda(1-x^2)\theta^{(1)}&=0~,
 \\
 \cos^2\!z\,\partial_z\big(\lambda x\theta^{(1)}+\tan z\,\rho^{(1)}\big)-\partial_xA_{\beta_2}^{(1)}-2\rho^{(1)}&=0~.
\end{align}
\end{subequations}
We change variables and define
\begin{align}\label{eqn:lin-transf}
 \phi&=\frac{\sec^2z }{x^2+\tan^2\!z} A_{\beta_2}^{(1)}~,
 &
 \zeta&=\frac{\lambda x \theta^{(1)}+\tan z\,\rho^{(1)}}{x^2+\tan^2\!z}~.
\end{align}
Eqns.~(\ref{eqn:lin-eq}) then become
\begin{subequations}\label{eqn:lin-eq4}
\begin{align}
 \cos^2\!z\partial_z\phi+(1-x^2)\partial_x\zeta&=0~,
 \\
 \partial_z\zeta-\partial_x\phi&=0~.
\end{align}
\end{subequations}
Taking a derivative of the first equation with respect to $x$ 
and then using the result in the second equation leaves us with
an equation for $\zeta$ alone,
\begin{align}
 \cos^2\!z\,\partial_z^2\zeta+\partial_x(1-x^2)\partial_x\zeta&=0~.
\end{align}
This equation can be solved by separation of variables. 
We write $\zeta=p_\ell(x)\zeta_\ell(z)$ and conveniently introduce a constant $\ell$ such that the above equation implies
\begin{align}
 \partial_x(1-x^2)\partial_x p&=-\ell(\ell+1)p~,
 &\cos^2\!z\,\partial_z^2\zeta_\ell&=\ell(\ell+1)\zeta_\ell~.
\end{align}
The $x$-dependent part is the Legendre equation, and the requirement of regularity at $x=1$ and $x=0$, 
corresponding to $\beta_1=0$ and $\beta_1=\pi$, respectively, 
forces us to choose the Legendre functions of the first kind, $P_\ell$, with $\ell$ an integer.\footnote{%
Note that regularity of $\theta$ and $\rho$ implies regularity of $\zeta$, $\phi$ for $z\in(0,\frac{\pi}{2})$.
}
Since $P_\ell=P_{-\ell-1}$, we can restrict to non-negative $\ell$ w.o.l.g.
The result should not come as a surprise, 
as the $P_\ell$ are the polynomials appearing in the spherical harmonics $Y_{\ell,0}$.
Summing up, we have
\begin{align}\label{eqn:zeta-separation}
 \zeta&=\sum_{\ell=0}^\infty P_\ell(x)\zeta_{\ell}(z)~,
 &
 \cos^2\!z\,\partial_z^2\zeta_\ell&=\ell(\ell+1)\zeta_\ell~.
\end{align}
Solving for $\zeta_\ell$ yields
\begin{align}\label{eq:zeta-tilde-sol}
 \zeta_\ell&=c_{\ell,1}f_{\ell}+c_{\ell,2}f_{-\ell-1}~,
 &
 f_\ell&=(\cos z)^{\ell+1} {}_2F_1\Big(\frac{\ell+1}{2},\frac{\ell+1}{2},\ell+\frac{3}{2},\cos^2\!z\Big)~.
\end{align}
Note that $\ell(\ell+1)$ is invariant under $\ell\rightarrow -\ell-1$, but $f_\ell$ is not.
The remaining task is to solve for $\phi$. 
We note that $\phi$ is defined in terms of the one form component $A_{\beta_1}$ by a simple rescaling.
In particular, it has to satisfy the same regularity conditions, namely vanish at $x=0$ and $x=1$, corresponding to $\beta_1=\pi$ and $\beta_1=0$, respectively.
We should therefore expand it in 1-form spherical harmonics.
Just like $A_{\beta_2}$, $\phi$ is part of a divergence-free vector field, 
and this fixes the kind of vector spherical harmonics that appear.
To make the normalization convention explicit, we take
\begin{align}\label{eq:one-form-Y}
\vec Y_{\ell,0}=\star^{}_{\mathrm{S}^2} d P_\ell=(1-x^2)\frac{d}{dx}P_{\ell}(x)~,
\end{align}
and expand
\begin{align}\label{eq:phi-vec-Y}
 \phi&=\sum_{\ell=0}^\infty \phi_\ell(z) \vec Y_{\ell,0}~.
\end{align}
Note that the $\ell=0$ term vanishes regardless of the value of $\phi_0$. 
Using this expansion in the second equation of (\ref{eqn:lin-eq4}) shows that
\begin{align}
 \zeta_0^\prime&=0~,
 &
 \phi_\ell&=-\frac{1}{\ell(\ell+1)}\zeta_\ell^\prime~,\quad \ell\geq 1~.
\end{align}
$\zeta_0^\prime$ is indeed a constant  and the first equation fixes $c_{0,1}=0$.
It will be convenient to recast the expansion of $\phi$ in one-form spherical harmonics as expansion in the Legendre polynomials themselves.
Using $(1-x^2)\frac{d}{dx}P_\ell=-\ell(xP_\ell-P_{\ell-1})$ and $(2\ell+1)x P_\ell=(\ell+1)P_\ell+\ell P_{\ell-1}$, we find
\begin{align}
 (1-x^2)\frac{d}{dx}P_\ell&=-\frac{\ell(\ell+1)}{2\ell+1}\left(P_{\ell+1}-P_{\ell-1}\right)~.
\end{align}
With the understanding that $\zeta_{-1}^\prime=\zeta_0^\prime=0$, the expansion for $\phi$ becomes
\begin{align}\label{eqn:phisol}
 \phi&=\sum_{\ell=0}^\infty P_\ell(x) \varphi_{\ell}(z)~,
 &
 \varphi_\ell&=\frac{\zeta_{\ell-1}^\prime}{2l-1}
   -\frac{\zeta_{\ell+1}^\prime}{2l+3}~.
\end{align}
To translate back to $\theta^{(1)}$ and $\rho^{(1)}$ we use (\ref{eqn:lin-transf}), which can be solved to give
\begin{align}\label{eqn:lin-transf-back}
 \theta^{(1)}&=\lambda(\sin z\cos z\,\phi+x\zeta)~,&
 \rho^{(1)}&=\tan z\,\zeta-x\cos^2\!z\, \phi~.
\end{align}
This yields
\begin{align}
 \theta^{(1)}&=\sum_{\ell=0}^\infty P_\ell(x)\theta_\ell^{(1)}(z)
 &
 \theta_\ell^{(1)}&=\lambda\Big[\sin z\cos z\varphi_\ell+\frac{\ell \zeta_{\ell-1}}{2\ell-1}+\frac{(\ell+1)\zeta_{\ell+1}}{2\ell+3}\Big]~,
 \\
  \rho^{(1)}&=\sum_{\ell=0}^\infty P_\ell(x)\rho^{(1)}_\ell(z)
 &
 \rho_\ell^{(1)}&=\tan z \zeta_{\ell}-\cos^2\!z\,\Big[\frac{(\ell+1)\varphi_{\ell+1}}{2\ell+3}+\frac{\ell \varphi_{\ell-1}}{2\ell-1}\Big]~.
\end{align}
Finally, the solution for the gauge field (\ref{eq:solA3}) with (\ref{eqn:lin-transf}) becomes
\begin{align}\label{eq:solA4}
 A_{\beta_2}^{(1)}&=-(1-x^2)\cos^2\!z\,\phi+\phi+\mathcal A_0^{(1)}~.
\end{align}
Simply substituting the expressions for $\rho^{(1)}$ and $\theta^{(1)}$ would yield an expansion in scalar spherical harmonics,
which would be straightforward and possible but make regularity obscure.
The more natural expansion is again in terms of the 1-form spherical harmonics (\ref{eq:one-form-Y}). 
The second term can be brought into this form directly, using (\ref{eq:phi-vec-Y}).
The first term can be rearranged using the expansion (\ref{eqn:phisol}) and $(2n+1)P_n=\frac{d}{dx}(P_{n+1} - P_{n-1})$.
The result is
\begin{align}
 A_{\beta_2}^{(1)}&=\sum_{\ell=1}^\infty A^{(1)}_\ell(z)(1-x^2)\frac{d}{dx}P_\ell(x)~,
 &
 A^{(1)}_\ell&=\cos^2z\Big[\frac{\phi_{\ell+1}}{2\ell+3}-\frac{\phi_{\ell-1}}{2\ell-1}\Big]-\frac{\zeta_\ell^\prime}{\ell(\ell+1)}~,
\end{align}
where $\mathcal A_0^{(1)}$ has been fixed to ensure regularity at the poles of $\rm S^2$.
All functions appearing as radial profiles are hypergeometric functions with argument $\cos^2 z$.
They have the usual regular singular points for $\cos^2z=0$ and $\cos^2z=1$ and are regular for $\cos^2z\in(0,1)$.
Geometrically, $\cos^2z=0$ corresponds to the conformal boundaries at $r\rightarrow\pm \infty$, 
while $\cos^2z=1$ corresponds to $r=0$ where we need regular solutions.
For any choice of the constants the solutions are all regular at $z=0$, and there are therefore no constraints from regularity in the interior of AdS$_5$.

With the general linearized solution in hand we can now analyze the near-boundary behavior from which the one-point functions can be deduced straightforwardly.
The D5-branes intersect the conformal boundary at $r\rightarrow \pm\infty$, corresponding to $z\rightarrow\pm\frac{\pi}{2}$, and each of these limits corresponds to an AdS$_3$ defect.
To analyze the near-boundary behavior of the solutions we switch to Fefferman-Graham coordinates 
$r=\mp\log (u/2)$ or
$z=\pm(2\tan^{-1}(\frac{2}{u})-\frac{\pi}{2})$.
This transforms the $\rm AdS_4$ metric in (\ref{eqn:metric-AdS5}) to Fefferman-Graham gauge, where the terms in the near-boundary expansion of the bulk fields can be related directly to the sources and vacuum expectation values in the CFT.
We then find
\begin{subequations}\label{eq:ads-slicing-exp}
\begin{align}
 \theta_\ell^{(1)}&= \lambda u^{1-\ell}(d_{\ell,2}+\dots)+\lambda u^{2+\ell}(d_{\ell,1}+\dots)~,
 \\
 \pm\frac{\rho_\ell^{(1)}-\ell\mathcal A_\ell^{(1)}}{2\ell+1}&=\left(\frac{d_{\ell+1,2}}{\ell+1}+\dots\right)u^{-\ell-1} - \left(\frac{\ell+3}{(2\ell+3)(2\ell+5)}d_{\ell+1,1}+\dots\right)u^{4+\ell}~,
 \\
 \pm\frac{\rho_\ell^{(1)}+(\ell+1)\mathcal A_\ell^{(1)}}{2\ell+1}&=
 \left(\frac{2-\ell}{(2\ell-3)(2\ell-1)}d_{\ell-1,2}+\dots\right)u^{3-\ell} 
 + \left(\frac{d_{\ell-1,1}}{\ell}+\dots\right)u^\ell~,
\end{align}
\end{subequations}
where the upper/lower choice of the sign in the last two lines corresponds to the behavior at the $\rm AdS_3$ obtained as $r\rightarrow +\infty$/$r\rightarrow -\infty$.
The dots denote terms which vanish at $u=0$ and we have redefined the coefficients as
\begin{align}
 d_{\ell,1}&=\frac{\ell (\ell+1) }{4 \ell^2-1}c_{\ell-1,1}+c_{\ell+1,1}~,
 &
 d_{\ell,2}&=c_{\ell-1,2}+\frac{\ell (\ell+1) c_{\ell+1,2}}{(2\ell+1)(2\ell+3)}~.
\end{align}
The leading terms in the expansions (\ref{eq:ads-slicing-exp}) show that the slipping mode corresponds to an operator with $\Delta_+=2+\ell$ and $\Delta_-=1-\ell$.
The two linear combinations of bending mode and gauge field source 
one operator with $\Delta_+=\ell+4$ and $\Delta_-=-1-\ell$, and another one with $\Delta_+=\ell$ and $\Delta_-=3-\ell$. 
In standard quantization the $d_{\ell,2}$ are the sources and the $d_{\ell,1}$ parametrize the expectation values.
For the low-lying operators the roles are exchanged if alternative quantization is chosen.
The scaling dimensions match those found for a flat defect in \cite{DeWolfe:2001pq}.

At the linearized level each perturbation can be turned on independently,
and as discussed above there are no further regularity conditions from the interior of the bulk AdS.
So the $d_{\ell,1}$ and $d_{\ell,2}$ can be chosen independently, and we can in particular set all sources to zero and 
still dial the subleading terms in the asymptotic expansions (\ref{eq:ads-slicing-exp}).
With all sources vanishing there are no extra contributions from holographic renormalization and the subleading terms 
directly correspond to the one-point functions.
The expectation values of the operators sourced by slipping mode, bending mode and gauge field are linked, 
since they are all three proportional to just one set of constants given by the $d_{\ell,1}$.
This of is a result of supersymmetry.
But we nevertheless find a large ``moduli space'' of supersymmetric vacuum states, where the one-point functions can be chosen 
on one of the $\rm AdS_3$ making up the defect and are then fixed on the remaining $\rm AdS_3$.
We can make this very explicit for the special case of a pure mass deformation,
where the slipping mode only has a non-vanishing $\theta_0$ mode and the bending mode only has a non-vanishing $\rho_1$ mode.
The explicit solution is given by
\begin{align}\label{eqn:delta-m-linear-embedding}
 \theta^{(1)}&=P_0(x)\cos z(m\sin z+c\cos z)~,
 &
 \rho^{(1)}&=\lambda P_1(x) \cos z (c\sin z-m\cos z)~,
\end{align}
where $m$ parametrizes the mass deformation and $c$ the chiral condensate.
Both can be varied independently, so the expectation value on one of the $\rm AdS_3$ parts of the defect
is not fixed in terms of the source on that part.
But sources and expectation values are related between the two $\rm AdS_3$ parts obtained as $z\rightarrow \pm\frac{\pi}{2}$.

We will close this section with qualitative comments on the higher orders in the perturbative expansion.
Starting from the linearized solution corresponding to a pure mass deformation in (\ref{eqn:delta-m-linear-embedding}),
it is straightforward to then expand and solve the BPS equations to higher orders in the fluctuations.
The linear-order fluctuations do not source the fields at quadratic order, and it is consistent to set them to zero
\begin{align}
 \theta^{(2)}&=\rho^{(2)}=0~.
\end{align}
At cubic order in the mass deformation, 
the linear-order fluctuations $\theta^{(1)}_0$ and $\rho_1^{(1)}$ do appear as sources for $\theta_0^{(3)}$ and $\rho_1^{(3)}$ and those modes are non-vanishing.
Moreover, they also trigger higher spherical harmonics in $\theta$ and $\rho$.
Namely, $\theta_0^{(1)}$ and $\rho_1^{(1)}$ appear as sources in the equations for $\theta_2^{(3)}$ and $\rho_3^{(3)}$,
and consistent solutions consequently require non-vanishing $\theta_2$ and $\rho_3$ as well.
At cubic order the solution therefore takes the form
\begin{align}
\theta^{(3)}&=\theta_0^{(3)}(z)P_0(x)+ \theta_2^{(3)}(z)P_2(x)~,
&
\rho^{(3)}&=\rho_1^{(3)}(z)P_1(x)+ \rho_3^{(3)}(z)P_3(x)~.
\end{align}
The general solution for each of the higher modes $\theta^{(3)}_0$, $\theta^{(3)}_2$ and $\rho^{(3)}_1$, $\rho^{(3)}_3$ 
is a linear combination of a particular solution to the inhomogeneous equation with the general solution to the homogeneous equation.
This introduces four additional constants and the general solution is rather bulky.
However, physically we still want to describe a pure mass deformation, which means the terms in the near-boundary expansions of 
$\theta^{(3)}_2$ and $\rho^{(3)}_3$ that correspond to sources in the CFT should be zero -- on both $\rm AdS_3$ parts that make up the defect.
This fixes two of the four constants, and afterwards the solutions for $\theta^{(3)}_2$ and $\rho^{(3)}_3$ take a very simple form, namely
\begin{align}\label{eq:mass-cubic-higher-harmonics}
 \theta^{(3)}_2&=-\frac{1}{3}\theta^{(1)}_0\left(\rho^{(1)}_1\right)^2~,
 &
 \rho^{(3)}_3&=\frac{1}{15}\left(\rho^{(1)}_1\right)^3~.
\end{align}
Expanding these modes near the conformal boundaries at $z\rightarrow\pm \frac{\pi}{2}$ shows that the terms corresponding to the expectation value of the dual operator in $\theta^{(3)}_2$ are non-zero, and the same applies for the two combinations of $\rho^{(3)}_3$ and $A_{3}^{(3)}$.
The remaining two constants appear in $\rho^{(3)}_1$ and $\theta^{(3)}_0$ only and they can be fixed by demanding that the mass and expectation values for the lowest spherical harmonics are not redefined at cubic order.
We will not give the explicit form for $\rho^{(3)}_1$ and $\theta^{(3)}_0$ here, and leave a more detailed study of the non-linear equations for the future.
At the technical level, the need for higher spherical harmonics implies that a separation ansatz can not be used to solve the BPS equations, at least not in these variables.
But the simple form of the higher modes in (\ref{eq:mass-cubic-higher-harmonics}) certainly suggests that a separation of variables ansatz may work after suitably chosen field redefinitions.

\section{\texorpdfstring{$\rm S^3$}{S3} defects and \texorpdfstring{$\rm S^4$}{S4} partition function from holography}\label{sec:S4-holographic}

In this section we analytically continue the embeddings for the AdS$_4$ slicing to global Euclidean AdS$_5$, where the slices are $\rm S^4$.
The role of IR regularity conditions is qualitatively different in that case and fixing the CFT sources fixes a unique vacuum state and a unique D5-brane embedding.
We study the perturbative embeddings for a pure mass deformation and compute the partition function for $\N{4}$ SYM on S$^4$ with massive 
defects on an equatorial $\rm S^3$ holographically, to quadratic order in the mass parameter.

To get from the Lorentzian AdS slicing to Euclidean spheres, we use the following strategy.
Firstly, the equations (\ref{eq:BPS}) do not depend on the chosen coordinates on the AdS$_3$ slices.
So we can straightforwardly implement the analytic continuation
\begin{align}
\label{eq:acont1}
r&=\hat r+\frac{i\pi}{2}~,
 &
 g_{\mathrm{AdS}_3}&\rightarrow -g_{\mathrm{S}^3}~.
\end{align}
The continuation from $\rm AdS_3$ to $\rm S^3$ can proceed via the continuation from $\rm AdS_3$ to $\rm dS_3$, as used in \cite{Karch:2015vra},
and from there to $\rm S^3$.
Note that this results in an $\rm S^3$ with negative signature.
This turns the AdS$_5$ part of the metric (\ref{eqn:metric-AdS5}) into
\begin{align}\label{eq:metric-S3-slice}
 g_{\mathrm{AdS}_5}&=d\rho^2+\cosh^2\!\rho\Big[d\hat r^2+\sinh^2\!\hat r\, g_{\mathrm{S}^3}\Big]~,
\end{align}
which is Euclidean AdS$_5$, or $\mathds{H}_5$, sliced by $\mathds{H}_4$ surfaces.
The massless embedding $\rho(r)\equiv \rho(\hat r)\equiv 0$ now describes a D5 wrapping $\mathds{H}_4\times$S$^2$,
and the deformations described by (\ref{eq:BPS}) with $r\rightarrow \hat r$ generally preserve the S$^3$ isometries.

We now implement a coordinate transformation such that the metric in (\ref{eq:metric-S3-slice}) turns into
\begin{align}
 g_{\mathrm{AdS}_5}&=dR^2+\sinh^2\!R\Big[d\chi^2+\cos^2\!\chi\, g_{\mathrm{S}^3}\Big]~,
\end{align}
where the terms in square brackets combine to an S$^4$.
This can be achieved by setting
\begin{align}\label{eqn:coord-transf-sphere}
 \cosh R&=\cosh\rho\cosh \hat r~,
 &
 \sinh R\sin\chi&=\sinh\rho~.
\end{align}
The embedding $\rho\equiv 0$ translates to $\chi\equiv 0$, which wraps an equatorial S$^3$ inside S$^4$, as desired.
As Euclidean DBI action for the D5-branes we have
\begin{align}\label{eqn:D5-action-E}
 S_{\mathrm{D5}}&=T_5\int_{\Sigma_6}d^6\xi \sqrt{\det\left(g+F\right)}+ T_5\int_{\Sigma_6}C_4\wedge F~,
\end{align}
with $C_4$ of the background solution given by
\begin{align}\label{eqn:RR-gauge-field-E}
 C_4&=\zeta(\chi)\sinh^4\!R\,\sin^2\!\alpha_1\sin\alpha_2\, dR\wedge d\alpha_1\wedge d\alpha_2\wedge d\alpha_3+\dots~,&\zeta^\prime(\chi)=-4\cos^3\chi~.
\end{align}
Note the sign in $\zeta^\prime$ to account for the ordering of $dR$ and $d\chi$ in $dC_4$, to get the positive volume form.

The BPS equations can be obtained by analytically continuing and coordinate transforming those for the $\rm AdS_4$ slicing,
and we can readily obtain the solution for the gauge field by that procedure from the solution for the $\rm AdS_4$ slicing in (\ref{eq:solA}).
This yields
\begin{align}\label{eq:solA-S}
 A_{\beta_2}&=i\lambda \sin\theta\cosh R-x \cos\theta\sinh R\sin\chi+\mathcal A_0~.
\end{align}

\subsection{Continuing perturbative solutions}
We now want to use (\ref{eqn:coord-transf-sphere}) to obtain perturbative embeddings for the sphere slicing.
For an arbitrary given embedding in the $\rm AdS_4$ slicing coordinates, described by $\rho$, $\theta$, $A_{\beta_2}$ as functions of $r$ and $\beta_1$, 
we first implement (\ref{eq:acont1}) and replace $r$ by $\hat r$.
We then use the perturbative expansion as set up in eq.~(\ref{eq:pert-exp})
and solve eq.~(\ref{eqn:coord-transf-sphere}) for $\hat r$ in terms of $R$ in an expansion in $\kappa$,
and likewise for $\chi$ in terms of $\rho$.
At linear order in the fluctuations this yields
\begin{align}
 r(R)&=R+\frac{i\pi}{2}+\mathcal O(\kappa^2)~,
 &
 \sinh R\,\chi^{(1)}(R)&=\rho^{(1)}(r^{(0)}(R))~.
\end{align}
To find the solution for the embedding in the $\rm S^4$ slicing, we analytically continue the equation, rather than the solution.
The linearized BPS equations for the $\rm S^4$ slicing with the analytic continuation derived above are given by (\ref{eqn:lin-eq}) with
$2\tanh^{-1}\tan(z/2)=r^{(0)}(R)$ and $\rho^{(1)}=-i\sec z\,\chi^{(1)}$.
The solution corresponding to a mass deformation is
\begin{subequations}\label{eq:linsol-S4}
\begin{align}
\theta^{(1)}&=\csch^2\!R\,(m \cosh R+c)~,
\qquad
\chi^{(1)}=i\lambda  \cos \beta_1 \csch^3\!R\,(m+c \cosh R)~.
\end{align}
For the gauge field the solution is obtained from (\ref{eq:solA-S}), 
and the constant $\mathcal A_0$ has to be fixed such that $A^{(1)}_{\beta_2}$ vanishes at $\beta_1=0$ and $\beta_1=\pi$, to get a regular one form on S$^2$.
This yields $\mathcal A_0=-i\kappa\lambda m$ and consequently
\begin{align}
A^{(1)}_{\beta_2}&=i \lambda  \sin ^2\!\beta_1 \csch^2\!R\, (c \cosh R+m)~.
\end{align}
\end{subequations}
Note that the gauge field and bending mode are imaginary.
For the gauge field this is parallel to the $\rm S^4$ slicing solutions for D7-branes obtained in \cite{Karch:2015vra},
and for the bending mode it naturally aligns with its mixing with the gauge field. 
The combined solution to the BPS equations in (\ref{eq:linsol-S4}) indeed solves the equations of motion derived from 
the DBI action (\ref{eqn:D5-action-E}), (\ref{eqn:RR-gauge-field-E}).

The next step is to implement regularity conditions at $R=0$. 
The expansion of the slipping and bending modes reads
\begin{align}
 \theta^{(1)}&=\frac{c+m}{R^2}+\mathcal O\left(1\right)~,
 &
 \sec\beta_1\chi^{(1)}&=\frac{i \lambda   (c+m)}{R^3}-\frac{i \lambda  m }{2 R}+\mathcal O\left(R\right)~.
\end{align}
The slipping mode can be rendered finite at the origin by setting 
\begin{align}
\label{eq:linreg}
 c&=-m~.
\end{align}
This leaves an $R^{-1}$ divergence in the bending mode. 
However, the relevant quantity that needs to be finite is $\sinh R\, \chi$ and since $\sinh R$ vanishes as $R\rightarrow 0$ this combination is indeed finite.
Eventually the bending mode mixes with the gauge field and we will see below that the decoupled modes are indeed regular,
so this completes the discussion of the regularity conditions.

\subsection{Holographic one-point functions}

In this section we compute the one-point functions for the operators sourced by the perturbations to the massless embedding discussed in the previous section.
The fluctuations satisfy a coupled set of equations of motion and we first need to reformulate the action in terms of linear combinations of $\chi$, $\theta$ and $A_{\beta_2}$ such that the equations of motion decouple.
It is these linear combinations that have a consistent near-boundary behavior with two generalized boundary values corresponding to source and expectation value for a dual operator.
We again work perturbatively and expand the action as
\begin{align}
 S_{\rm D5}&=\sum_n \kappa^n S^{(n)}_{\rm D5}~.
\end{align}
The $\mathcal O(\kappa^0)$ part is independent of the fluctuations but will nevertheless be useful for the computation of the partition function.
It is given by
\begin{align}\label{eq:sphere-action0}
 S^{(0)}_{\rm D5}&=T_5V_{\rm S^2} V_{\rm S^3}\int d R\sinh^3\!R~.
\end{align}
The interesting part for the fluctuations is the $\mathcal O(\kappa^2)$ term, which is given by
\begin{align}\label{eq:sphere-action}
\frac{2S^{(2)}_{\rm D5}}{V_{\rm S^1} V_{\rm S^3}T_5} =\int d Rd\beta_1\sin\beta_1 \sinh^3\!R\Bigg[
&
(\partial_{\beta_1}\theta^{(1)})^2
+
(\partial_{ R}\theta^{(1)})^2-2(\theta^{(1)})^2
\nonumber\\ &
+\sinh^2 R((\partial_{\beta_1}\chi^{(1)})^2+(\partial_{ R}\chi^{(1)})^2)-3(\chi^{(1)})^2
\nonumber\\ &
+\frac{(\partial_{\beta_1}A_{\beta_2}^{(1)})^2+(\partial_{ R}A_{\beta_2}^{(1)})^2}{\sin^2\beta_1}
-\frac{8\sinh R}{\sin\beta_1}\,\chi^{(1)}\partial_{\beta_1}A_{\beta_2}^{(1)}
 \Bigg]~.
\end{align}
Gauge field and bending mode are coupled, and to decouple them we define
\begin{subequations}
\begin{align}
 \varpi^{(1)} &= \csc^2\!\beta_1 A_{\beta_2}^{(1)}-\sec\beta_1\sinh R\, \chi^{(1)}~,
 \\
 \varsigma^{(1)} &= 2\csc^2\!\beta_1 A_{\beta_2}^{(1)} + \sec\beta_1\sinh R\, \chi^{(1)}~.
\end{align}
\end{subequations}
The resulting action, using that we have $\ell=1$ modes for $\rho$ and $A_{\beta_2}$ and an $\ell=0$ mode for $\theta$,
and after integrating over $\beta_1$, can be written conveniently as
\begin{align}
 S^{(2)}_{\rm D5}&=T_5 V_{\rm S^1}V_{\rm S^3}\left(S_{\theta^{(1)}}+S_{\varsigma^{(1)}}+S_{\varpi^{(1)}}\right)~,
\end{align}
where
\begin{subequations}
\begin{align}
 S_{\theta^{(1)}}&=\int dR \sinh^3\!R\left((\partial_R\theta^{(1)})^2-2(\theta^{(1)})^2\right)~,
 \\
 S_{\varsigma^{(1)}}&=\int dR \sinh^3\!R\frac{1}{54}\left[
 6(\partial_R\varsigma^{(1)})^2-4\varsigma^{(1)}\coth R\,\partial_R \varsigma^{(1)}+(\varsigma^{(1)})^2\csch^2\!R(5-9\cosh(2R))
 \right]~,
 \\
 S_{\varpi^{(1)}}&=\int dR \sinh^3\!R\frac{2}{27}\left[
  3(\partial_R\varpi^{(1)})^2-4\varpi^{(1)}\coth R\,\partial_R \varpi^{(1)}-4(\varpi^{(1)})^2(\csch^2\!R-6)
 \right]~.
\end{align}
\end{subequations}
The corresponding equations of motion are
 \begin{align}\label{eq:decoupled-linear}
 \left(D_R^2 -10\right)\varpi^{(1)}&=0~,
 &
 \left(D_R^2 + 2\right)\varsigma_1^{(1)}&=0~,
 &
 \left(D_R^2 + 2\right)\theta^{(1)}&=0~,
 \end{align}
where $D_R^2=\csch^3\!R\,\partial_R \sinh^3\!R\,\partial_R$.
This matches the linearized spectrum for the coupled sector analysis found in \cite{DeWolfe:2001pq} for $\ell=1$.

For the solutions in (\ref{eq:linsol-S4}) the combination $\varpi^{(1)}$ vanishes identically, $\varpi^{(1)}\equiv 0$.
For the remaining two fields we will need the near-boundary expansions.
Transforming to Fefferman-Graham coordinates by setting $R=-\log(\epsilon/2)$ and expanding in small $\epsilon$ yields
\begin{align}\label{eq:asympt-exp}
 \theta^{(1)} &=m\epsilon+c\epsilon^2+\mathcal O(\epsilon^3)~,
 &
 \varsigma^{(1)} &=3i\lambda c\epsilon+3i\lambda m \epsilon^2+\mathcal O(\epsilon^3)~.
\end{align}
We note that $\varsigma^{(1)}$ is indeed regular at $R=0$ with the relation between $m$ and $c$ in (\ref{eq:linreg}).

\subsubsection{Holographic renormalization}
Since $\varpi^{(1)}\equiv 0$ we only need to discuss the two remaining fields.
A crucial point to notice is that $\varsigma^{(1)}$ is a scalar with alternative quantization, which requires a Legendre transformation in addition to the usual holographic counterterms.
Moreover, the action is not in the standard form for a Klein-Gordon field, but differs by total derivatives.
So we will discuss the holographic renormalization in some detail.
The variation of the action evaluated on shell reads
\begin{subequations}
\begin{align}
 \delta S_{\theta^{(1)}}&=2\sinh^3\!R\delta\theta\partial_R\theta\Big\vert_{R\rightarrow\infty}
 =-\frac{2}{\epsilon}\delta\theta^{(1)}_0 \theta^{(1)}_0 -2(\theta^{(1)}_0\delta\theta^{(1)}_1+2\theta^{(1)}_1\delta\theta^{(1)}_0)+\mathcal O(\eps)~,
 \\
  \delta S_{\varsigma^{(1)}}&=\frac{2}{27}\sinh^3\!R\,\delta\varsigma^{(1)}\left(3\partial_R-\coth R\right)\varsigma^{(1)}\Big\vert_{R\rightarrow\infty}
 \nonumber\\
 &=
 -\frac{8}{27\eps}\varsigma^{(1)}_0\delta \varsigma^{(1)}_0-\frac{2}{27}\left(4\varsigma^{(1)}_0\delta \varsigma^{(1)}_1+7\varsigma^{(1)}_1\delta \varsigma^{(1)}_0\right)+\mathcal O(\eps)~,
\end{align}
\end{subequations}
For the near-boundary expansions we have used $R=-\log(\epsilon/2)$ and $\partial_R=-\epsilon\partial_\eps$.
Moreover, with a slight abuse of notation, we expanded $\theta^{(1)}=\epsilon \theta_0^{(1)}+\epsilon^2\theta_1^{(1)}+\dots$ and used an analogous expansion for $\delta\theta^{(1)}$,
and likewise for $\varsigma^{(1)}$ and $\delta \varsigma^{(1)}$.
We will not use an explicit subscript on $\theta^{(1)}$ to denote the harmonic on $\rm S^2$, such that no confusion should arise.
We add the holographic counterterms
\begin{subequations}\label{eqn:ct}
\begin{align}
 S_{\theta^{(1)},\rm ct}&=\int dR \sinh^3\!R \,(\theta^{(1)})^2~,
 \\
 S_{\varsigma^{(1)},\rm ct}&=-\frac{2}{27}\int dR \sinh^3\!R \,\left[(\varsigma^{(1)})^2+3\varsigma^{(1)}\partial_R\varsigma^{(1)}\right]~.
\end{align}
\end{subequations}
This produces, as desired, a valid variation for alternative quantization for $\varsigma$.
The variations and correspondingly the one-point functions of the CFT operators sourced by $\theta^{(1)}$ and $\varsigma^{(1)}$ are given by
\begin{subequations}\label{eq:deltaS}
\begin{align}
 \delta\left(S_{\theta^{(1)}}+S_{\theta^{(1)},\rm ct}\right)&=-2\theta^{(1)}_1\delta\theta^{(1)}_0=\langle \mathcal O_\theta\rangle \delta\theta^{(1)}_0~,
 \\
 \delta\left(S_{\varsigma^{(1)}}+S_{\varsigma^{(1)},\rm ct}\right)&=\frac{2}{9}\varsigma^{(1)}_0\delta \varsigma^{(1)}_1=\langle \mathcal O_{\varsigma}\rangle \delta \varsigma^{(1)}_1~.
\end{align}
\end{subequations}
We note that the counterterms in (\ref{eqn:ct}), to quadratic order in the fluctuations, are unique. 
With the scaling dimensions of the fields one may in principle add finite counterterms like $(\theta^{(1)})^3$ and $(\varsigma^{(1)})^3$ with arbitrary coefficients, 
and such counterterms can indeed be crucial \cite{Freedman:2016yue}.
However. such counterterms would only contribute starting at cubic order in the fluctuations, and do not affect the results derived here.
With the near-boundary expansion in (\ref{eq:asympt-exp}) we thus find
\begin{align}\label{eq:one-pt-S4}
 \langle \mathcal O_\theta\rangle &=-2c~,
 &
 \langle \mathcal O_\varsigma\rangle&=\frac{2}{3}i\lambda c~.
\end{align}
With the regularity relation in (\ref{eq:linreg}) we see that the one-point functions are completely fixed in terms of the source, as expected for the field theory on $\rm S^4$.
This is to be contrasted to the $\rm AdS_4$ slicing discussed in sec.~\ref{sec:general-linearized}, where the one-point functions could be dialed independently.

\subsection{Sphere partition function}
We now compute the contribution of the defect fields to the  $\rm S^4$ partition function, $\mathcal F(\rm S^4)$, to quadratic order in the mass deformation.
The partition function is given by $\mathcal F(\rm S^4)=-S_{\rm on-shell, ren}$, 
where $S_{\rm on-shell, ren}$ is the combined on-shell action for type IIB supergravity and the DBI action describing the D5 branes.
In the probe limit the contribution from the D5 branes is given by
\begin{align}
\delta\mathcal F(\rm S^4)&=-S_{\rm D5,\rm ren}~,
\end{align}
where $S_{\rm D5,\rm ren}$ is the renormalized on-shell D5-brane action.
The task at hand therefore is to compute $S_{\rm D5,\rm ren}$ to quadratic order in $m$.

We can straightforwardly compute the contribution at $\mathcal O(m^0)$, 
which is well defined since there are no finite counterterms that could be added on the holographic side.
This proceeds by directly integrating (\ref{eq:sphere-action0}) and taking the finite part -- the divergences are cancelled by the appropriate holographic counterterms without affecting the finite part -- which yields
\begin{align}
 S_{\rm D5,\rm ren}\big\vert_{m=0}&=\frac{4}{3}T_5V_{\rm S^1}V_{\rm S^3}~.
\end{align}
The term quadratic in $m$ can be conveniently computed using
\begin{align}
 \frac{\partial S_{\rm D5}}{\partial m}=
 \kappa^2\left[
 \frac{\delta S^{(2)}_{\rm D5}}{\delta \theta^{(1)}_0}\frac{\delta \theta^{(1)}_0}{\delta m}+
 \frac{\delta S^{(2)}_{\rm D5}}{\delta \varsigma^{(1)}_1}\frac{\delta \varsigma^{(1)}_1}{\delta m}
 \right]
 &=
 T_5 V_{\rm S^1}V_{\rm S^3}\left[
 \langle\mathcal O_\theta\rangle+3i\lambda \langle \mathcal O_\varsigma\rangle
 \right]
 \nonumber\\&
 =-4c \, T_5 V_{\rm S^1}V_{\rm S^3}~,
\end{align}
where (\ref{eq:deltaS}) and (\ref{eq:one-pt-S4}) were used for the second and third equalities.
We thus find, with the regularity relation (\ref{eq:linreg}), 
\begin{align}
 S_{\rm D5,\rm ren}&= T_5V_{\rm S^1}V_{\rm S^3}\left[\frac{4}{3} + 2\kappa^2m^2+\mathcal O(\kappa^3m^3)\right]~.
\end{align}
To identify the parameters on the holographic side with those on the field theory side, we first 
introduce $T_0$ defined by $T_{\rm D5}V_{\rm S^2}=T_0$, in analogy to the D3/D7 case in \cite{Karch:2015kfa}.
We then use $T_0= N_fN_c\sqrt{\lambda}/(2\pi^3)$ from the table above eq.~(4.18) in \cite{Chang:2013mca}.
This yields
\begin{align}
 T_{\rm D5}&=\frac{\mu}{4\pi^3} N_f N_c~,
 &
 \mu&=\frac{\sqrt{\lambda}}{2\pi}~.
\end{align}
With $V_{\rm S^1}=2\pi$ and $V_{\rm S^3}=2\pi^2$, we thus have $T_{\rm D5}V_{\rm S^1}V_{\rm S^3}=\mu N_f N_c$.
Finally, we note that the mass of the fundamental fields on the field theory side is given in terms of the leading term in the near-boundary expansion of the slipping mode by $M=\kappa m \mu$ \cite{Kruczenski:2003be}.
Consequently, the final result for the defect contribution to the partition function reads
\begin{align}\label{eq:F-holo}
 \delta\mathcal F(\rm S^4)&=- \mu N_f N_c\left[\frac{4}{3} + \frac{2M^2}{\mu^2}+\dots\right]~,
\end{align}
where the dots denote subleading terms in the mass and strong coupling expansions.

\section{\texorpdfstring{$\rm S^4$}{S4} partition function from supersymmetric localization}\label{sec:localization}

In this section we switch to the field theory side and compute the contribution of the defect fields to the partition function on $\rm S^4$ using supersymmetric localization.
We will start by discussing the matrix model that arises from the localization procedure and then evaluate
the partition function in the quenched approximation where the number of defect fields is small compared to 
the rank of the gauge group in the background \N{4} SYM theory.
This result can then be compared to the holographic computation of the same quantity in the previous section.

\subsection{The matrix model}

To fix notation we start with a brief review of supersymmetric localization.
For a gauge theory described by a partition function $\mathcal Z$, one
starts with the identification of a linear combination of a supercharge generating a supersymmetry that closes off shell
and a BRST charge, $\mathcal Q\equiv Q_{\rm SUSY} + Q_{\rm BRST}$.
The action is then deformed by a term $\mathcal Q V$ with appropriately chosen $V$ and this leaves the theory unchanged.
That is, instead of the original partition one evaluates
\begin{align}
\mathcal Z(t) &= \int DX~ e^{-S[X] - t\mathcal Q V[X]}~.
\end{align}
The original form of the partition function is recovered for $t=0$.
But since $\partial_t \mathcal Z(t)=0$, we can equivalently evaluate it for $t\rightarrow\infty$, where the integral localizes to the saddle points of $\mathcal Q V$.

For SU($N_c$) \N{4} SYM with conveniently chosen $V$, the saddles are parametrized by a single adjoint matrix, and the resulting expression for the partition function on $\rm S^4$ is \cite{Pestun:2007rz}
\begin{align}
\mathcal Z&=\int\! da^{N_c-1}\prod_{i<j} a_{[ij]}^2\, e^{S^{}_0}\,,&
S_0&={-\frac{8\pi^2}{\lambda}N_c\sum_i a_i^2}\,,
\end{align}
where $a_{[ij]}=a_i-a_j$ labels the roots of $\mathfrak{su}(N_c)$ and $a_i$ labels the weights.
Notably, the contributions of the one-loop fluctuations of the \N{4} SYM fields around the saddles of $\mathcal Q V$ cancel out and do not appear in this expression.
For \N{4} SYM coupled to fundamental fields, the saddles of $\mathcal Q V$ are still parametrized by a single adjoint matrix \cite{Pestun:2007rz},
and the only change in the partition function is the contribution of the 1-loop fluctuations of the fundamental fields around the saddles.
We used this result in \cite{Karch:2015kfa} to compare the holographic computation of the partition function to a field theory computation.

This story could potentially change when deforming \N{4} SU$(N_c)$ SYM theory on an S$^4$ by adding a defect matter sector.
We can no longer rely on the matrix model constructed out of the components listed in \cite{Pestun:2007rz}.
Generally, we would start from scratch to derive the new localized theory and compute the one-loop determinants.
However, inspired by the way the computation proceeds for gauge theories coupled to spacetime-filling matter fields, we will take a more pragmatic route here as follows.
The form of the one-loop determinant for $N_f$ (massive) fundamental hypermultiplets on an S$^3$, coupled to an intrinsically 3-dimensional superconformal Chern-Simons theories has been given in \cite{Kapustin:2009kz} and reads
\begin{align}
\prod_{i} \sech^{N_f}\big(\pi(a_i +M)\big)~,
\end{align}
where $M$ is the mass parameter.
We will assume that, with appropriately chosen $V$, the path integral for the defect theory we are considering once again localizes to saddle points parametrized by the single adjoint matrix setting the constant value for one of the scalar fields in the \N{4} SYM multiplet.
The only contribution of the defect fields would then again be their one-loop fluctuations, and the matrix model we have to solve is
\begin{align}\label{eq:defect-Z}
 \mathcal Z_{\rm defect}&=
 \int da^{N_c-1}\prod_{i<j} a_{[ij]}^2\,\frac{1}{\prod_i \cosh^{N_f}\big(\pi(a_i + M)\big)}~e^{S_0}~.
\end{align}
This procedure for constructing the matrix model may be further motivated by the observation that
the localizing supercharge $\mathcal Q$ and the cohomological deformation $\mathcal Q V$ used for the three-dimensional theories in \cite{Kapustin:2009kz} 
are the same as those used for the four-dimensional theory in \cite{Pestun:2007rz}.
This was pointed out in appendix B of \cite{Drukker:2010jp}.

\subsection{\texorpdfstring{$\rm S^4$}{S4} partition function}

To evaluate the partition function (\ref{eq:defect-Z}) in the limit of strong coupling and large $N_c$, with $N_f\ll N_c$, we first rearrange the integrand such that all factors appear in the exponent. 
This yields, with $\zeta=N_f/N_c$,
\begin{align}
  \mathcal Z_{\rm defect}&=
 \int da^{N_c-1}\,e^{S}~,
 &
 S&=S_0+\sum_{i<j}\log a_{[ij]}^2 - N_c\zeta\sum_i\log \big(\cosh(\pi(a_i + M)\big)~.
\end{align}
The first two terms in $S$ are $\mathcal O(N_c^2)$, while the last term is $\mathcal O(N_f N_c)$ and provides the contribution of the defect fields.
In the quenched approximation with $N_f\,{\ll}\, N_c$ 
the computation of the partition function  can be organized in an expansion in $\zeta$,
by using that we are close to the solution for the matrix model for pure \N{4} SYM.
The expansion of the ``on-shell'' action reads
\begin{align}
\frac{S}{N_c^2} &= \tilde{S}_{0}|_{\rho_0} + \zeta(S_1|_{\rho_0}+\delta \tilde{S}_0|_{\rho_0})+\CO(\zeta^2)~,
\end{align}
where we have defined $\tilde S_0$ by $N_c^2\tilde{S}_0=S_0+\sum_{i<j}\log a_{[ij]}^2$ 
and $S_1$ is the contribution of the defect fields.
We have denoted by $\rho_0$ the solution to the Gaussian matrix model corresponding to pure \N{4} SYM in the continuum limit,
which is the Wigner semicircle distribution
\begin{align}
\rho_0 (x)&= \frac{2}{\pi\mu^2}\sqrt{\mu^2-x^2}~,
&
\mu &= \frac{\sqrt{\lambda}}{2\pi}~.
\end{align}
The great simplification in the quenched approximation is that, since $\rho_0$ extremizes $\tilde S_0$, 
$\delta \tilde{S}_0|_{\rho_0}=0$ and we only need $S_1\vert_{\rho_0}$
to compute the contribution of the defect fields to the partition function at leading order.
The contribution of the defect fields thus evaluates to
\begin{align}
 \delta \mathcal F&=N_c^2\zeta S_1\vert_{\rho_0}~,
 &
 S_1\vert_{\rho_0}&=
 -\int_{-\mu}^\mu dx \rho_0 (x) \log\cosh\left(\pi(x+M)\right)~.
\end{align}
With a simple change of variables, setting $x=\mu y$, and defining $m$ by $M=\mu m$, we can rewrite this as
\begin{align}
\delta \mathcal F&=-\frac{2N_c^2 \zeta}{\pi}\int_{-1}^1 dy \sqrt{1-y^2} \log\cosh\left(\mu\pi(y+m)\right)~.
\end{align}
In analogy to the holographic computation we will evaluate this integral in an expansion for small $m$.
The $m=0$ contribution can be evaluated straightforwardly, yielding
\begin{align}
 \delta \mathcal F \big\vert_{m=0}&=-\frac{4N_c^2 \zeta}{\pi}\int_{0}^1 dy \sqrt{1-y^2} \log\cosh\left(\pi\mu y\right)\nonumber\\
 &\approx-\frac{4\mu N_c^2 \zeta}{\pi}\int_{0}^1 dy \sqrt{1-y^2} \pi\mu y=-\frac{4}{3}\mu N_c^2 \zeta~,
\end{align}
where we used $\mu\gg1$ and the large argument expansion for the $\cosh$ in the second line. 
This is justified except for a region with width of order $1/\mu$ around the origin. 
But since the integrand is smooth there the contribution from this region is subleading in $\mu$.

To compute the first term in the Taylor expansion of $\delta F$ around $m=0$, we evaluate
\begin{align}
 \frac{d\delta \mathcal F}{dm}\Big\vert_{m=0}&=
 -2\mu N_c^2\zeta\int_{-1}^1 dy\sqrt{1-y^2}\tanh(\pi\mu y)~.
\end{align}
The integrand is odd under $y\rightarrow -y$ and the integration domain symmetric, so $d\delta \mathcal F/dm\vert_{m=0}$ vanishes.
The next term in the expansion can be computed from
\begin{align}
 \frac{d^2 \delta \mathcal F}{dm^2}\Big\vert_{m=0}&=-2\pi\mu^2 N_c^2\zeta\int_{-1}^1 dy\sqrt{1-y^2}\sech^2(\pi\mu y)~.
\end{align}
For large $\mu$ the $\sech^2(\pi\mu y)$ factor decays exponentially away from $y=0$, so the dominant contribution comes from an exponentially small region around $y=0$.
At small $y$, we can expand $\sqrt{1-y^2}\approx 1-y^2/2$,
and since the contributions from regions where $y$ is $\mathcal O(1)$ are exponentially suppressed, we can use this expansion for the entire domain of integration.
As a result, to leading order in large $\mu$,
\begin{align}
 \frac{d^2 \delta \mathcal F}{dm^2}\Big\vert_{m=0}&=
 -2\pi\mu^2 N_c^2\zeta\int_{-1}^1 dy\sech^2(\pi\mu y)
 =-4\mu N_c^2\zeta~.
\end{align}
The explicit form of the contribution of the defect fields to the partition function, to quadratic order in $m$,  therefore reads
\begin{align}
 \delta \mathcal F&=\mu N_fN_c \left(-\frac{4}{3}  -2m^2+\dots\right)
 =-\mu N_f N_c\left(\frac{4}{3}  +\frac{2M^2}{\mu^2}+\dots\right)~.
\end{align}
This matches precisely the result of the holographic computation in (\ref{eq:F-holo}).
It validates not only the holographic computation but also lends support to the construction
of the matrix model outlined in the previous section and resulting in (\ref{eq:defect-Z}).
Extending the matrix model computation to higher orders in $m$ is straightforward, both in principle and in practice,
while the holographic computation gets considerably more involved at higher orders in $m$.
But already the computation thus far provides a strong check and we certainly expect the higher orders to agree as well.

\begin{acknowledgments}
We are very happy to thank Andreas Karch for many useful discussions.
The work of BR is supported, in part, by the U.S.~Department of Energy under Grant No.~DE-SC0011637. 
The work of CFU is supported, in part, by the U.S.~Department of Energy under Grant No.~DE-SC0011637 
and by the National Science Foundation under grant PHY-16-19926.
\end{acknowledgments}

\appendix

\section{Details on non-linear \texorpdfstring{$\kappa$}{kappa}-symmetry}\label{app:non-linear-kappa}

In this appendix we provide the details for the derivation of the non-linear $\kappa$-symmetry conditions discussed in sec.~\ref{sec:non-linear-kappa}.
The basic idea will be to isolate the $\rm S^5$ and $\rm AdS_5$ Clifford algebra structures and use the massive projector as well as the chirality constraint on the constant spinor $\epsilon_0$ to reduce to a minimal number of $\rm AdS_5$ Clifford algebra structures, which then have to vanish independently.
The starting point is the $\kappa$-symmetry condition as spelled out in (\ref{eqn:kappa-nonlinear1}), which we repeat for convenience
\begin{align}\label{eqn:kappa-nonlinear1-app}
 -\hat\Gamma e^{-i\rho\Gamma_{\underline{\rho}}\GammaAdS}e^{-i\theta\Gamma_{\underline{\theta}}\GammaS}\Gamma_{\underline{\rho}}\GammaUbeta\GammaAdS\epsilon
 +\frac{i}{2}\gamma^{ij}F_{ij}\hat\Gamma\epsilon&=h\epsilon~.
\end{align}
To evaluate $\hat\Gamma$ explicitly, we start from (\ref{eqn:Gammahat}), (\ref{eqn:gammarbeta}) and decompose
\begin{align}\label{eqn:decomp1}
 \gamma_{r\beta_1\beta_2}&=\mathcal A^{\mathds{1}}+\mathcal A^\rho\Gamma_\rho+\mathcal A^r\Gamma_r+A^{\rho r}\Gamma_{\rho r}~,
\end{align}
where, with $\tau^\m=\varepsilon^{\m \n \mathfrak{r}}(d\rho)_\n (d\theta)_{\mathfrak{r}}$ and $\varepsilon^{r\beta_1\beta_2}=1$,
the $\rm S^5$ Clifford algebra structures are given by
\begin{subequations}
\begin{align}
 \mathcal A^{\mathds{1}}&=(d\theta)_r\Gamma_{\theta}\Gamma_{\vec{\beta}}~,
 &
 \mathcal A^\rho&=
 (d\rho)_r\Gamma_{\vec{\beta}}-\tau^{\beta_i}\Gamma_{\beta_i}\Gamma_\theta~,
 \\
 A^r&=\Gamma_{\vec{\beta}}-\varepsilon^{r\m\n}(d\theta)_{\m}\Gamma_{\n}\Gamma_\theta~,
 &
 A^{\rho r}&=\varepsilon^{r\m \n}(d\rho)_{\m}\Gamma_{\n}-\tau^r \Gamma_\theta~.
\end{align}
\end{subequations}
With (\ref{eqn:decomp1}) and $\Gamma_{11}\epsilon=\epsilon$ we then find
\begin{align}
 -\hat\Gamma e^{-i\rho\Gamma_{\underline{\rho}}\GammaAdS}e^{-i\theta\Gamma_{\underline{\theta}}\GammaS}\Gamma_{\underline{\rho}}\GammaUbeta\GammaAdS\epsilon
 &=
 \Big[
 \mathcal M^{\mathds{1}}+\mathcal M^{\rho}\Gamma_{\rho}
 +\mathcal M^{r}\Gamma_{r}+\mathcal M^{\rho r}\Gamma_{\rho r}
 \Big]e^{-i\theta\Gamma_{\underline{\theta}}\GammaS}\GammaUbeta\epsilon~,
\end{align}
where
\begin{subequations}
\begin{align}
 \mathcal M^{\mathds{1}}&=-\cosh^2\!\rho\big(\mathcal A^r+i\tanh\rho A^{\rho r}\GammaS\big)~,
 &
 \mathcal M^{\rho}&=-\cosh^2\!\rho\big(A^{\rho r}-i\tanh\rho\mathcal A^r\GammaS\big)~,
 \\
 \mathcal M^{r}&=-\mathcal A^{\mathds{1}}-i\tanh\rho \mathcal A^\rho \GammaS~,
 &
 \mathcal M^{\rho r}&=-\mathcal A^\rho+i\tanh\rho \mathcal A^{\mathds{1}}\GammaS~.
\end{align}
\end{subequations}
Likewise, we decompose the $\gamma\cdot F$ term into $\rm AdS_5$ and $\rm S^5$ Clifford algebra structures and find
\begin{align}\label{eqn:decomp2}
 \frac{1}{2}\gamma^{ij}F_{ij}=\mathcal B^{\mathds{1}}+\mathcal B^\rho\Gamma_\rho+\mathcal B^r\Gamma_r+B^{\rho r}\Gamma_{\rho r}~,
\end{align}
with the $\rm S^5$ Clifford algebra structures
\begin{subequations}
\begin{align}
  B^{\mathds{1}}&=(d\theta)_r F^{r\beta_i}\Gamma_\theta \Gamma_{\beta_i}+
     F^{\beta_1\beta_2}\big(\Gamma_{\vec{\beta}}+\Gamma_\theta\varepsilon^{r\m \n}(d\theta)_{\m}\Gamma_{\n}\big)~,
  &
  B^r&=-F^{r\beta_i}\big(\Gamma_{\beta_i}+(d\theta)_{\beta_i}\Gamma_\theta\big)~,
  \\
  B^{\rho}&=F^{\m \n}(d\theta)_{\m}(d\rho)_{\n}\Gamma_\theta+F^{\beta_i \m}(d\rho)_\m \Gamma_{\beta_i}~,
  &
  B^{\rho r}&=-F^{r\beta_i}(d\rho)_{\beta_i}\mathds{1}~.
\end{align}
\end{subequations}
For the combination appearing in (\ref{eqn:kappa-nonlinear1-app}) this yields
\begin{align}
 \frac{i}{2}\gamma^{ij}F_{ij}\hat\Gamma\epsilon&=-i\left[\mathcal E^{\mathds{1}}+\mathcal E^\rho \Gamma_\rho+\mathcal E^r\Gamma_r+\mathcal E^{\rho e}\Gamma_{\rho r}\right]\GammaS\epsilon~,
\end{align}
where
\begin{subequations}
\begin{align}
 \mathcal E^{\mathds{1}}&=-\cosh\rho\big(\mathcal B^{\rho r}\mathcal A^{\mathds{1}}+\mathcal B^{\mathds{1}}\mathcal A^{\rho r}+\mathcal B^\rho\mathcal A^r-\mathcal B^r\mathcal A^\rho\big)~,
 \\
 \mathcal E^\rho&=-\cosh\rho\big(\mathcal B^r\mathcal A^{\mathds{1}}-\mathcal B^{\mathds{1}}\mathcal A^r+\mathcal B^\rho\mathcal A^{\rho r}+\mathcal B^{\rho r}\mathcal A^\rho\big)~,
 \\
 \mathcal E^r&=\sech\rho\big(\mathcal B^\rho\mathcal A^{\mathds{1}}-\mathcal B^{\mathds{1}}\mathcal A^\rho\big)-\cosh\rho\big(\mathcal B^r\mathcal A^{\rho r}+\mathcal B^{\rho r}\mathcal A^r\big)~,
 \\
 \mathcal E^{\rho r}&=\sech\rho\big(\mathcal B^{\mathds{1}}\mathcal A^{\mathds{1}}+\mathcal B^\rho\mathcal A^\rho\big)+\cosh\rho\big(\mathcal B^r\mathcal A^r-\mathcal B^{\rho r}\mathcal A^{\rho r}\big)~.
\end{align}
\end{subequations}
We can then assemble the full $\kappa$-symmetry condition, which becomes
\begin{align}\label{eqn:decomp3}
 \left[\mathcal N^{\mathds{1}}+\mathcal N^\rho \Gamma_{\rho}+\mathcal N^r \Gamma_r+\mathcal N^{\rho r} \Gamma_{\rho r}\right]\epsilon
 &=h\epsilon~,
\end{align}
where the $\rm S^5$ Clifford algebra structures are given by
\begin{align}
 \mathcal N^X&=\mathcal M^X e^{-i\theta\Gamma_{\underline{\theta}}\GammaS}\GammaUbeta -i\mathcal E^X\GammaS~,
 &
 X&\in\lbrace \mathds{1},\rho,r,\rho r\rbrace~.
\end{align}
We can now use the massive projector (\ref{eqn:massive-projector}) to reduce the number of Clifford algebra structures as follows.
Multiplying eq.~(\ref{eqn:decomp3}) by $\RAdS^{-1}\RS^{-1}$ yields
\begin{align}\label{eqn:kappa-non-linear-RA}
 \Big(
   \mathcal R_S[\mathcal N^{\mathds{1}}]
   +\mathcal R_S[\mathcal N^\rho]\mathcal R_A[\Gamma_\rho]
   +\mathcal R_S[\mathcal N^r]\mathcal R_A[\Gamma_r]
   +\mathcal R_S[\mathcal N^{\rho r}]\mathcal R_A[\Gamma_{\rho r}]
 \Big)\epsilon_0&=
 h\epsilon_0~,
\end{align}
where we use the definitions of $\mathcal R_S$ and $\mathcal R_A$ as given in (\ref{eqn:Rop}).
We now use the following identities
\begin{subequations}
\begin{align}
 \mathcal R_A[\Gamma_{\underline{\rho}}]&=\cosh r \,\mathcal K\, \Gamma_{\underline{\rho}}+i \sinh r \,\Gamma_{\underline{\rho r}}\GammaAdS~,
 \\
 \mathcal R_A[\Gamma_{\underline{r}}]&=
 \mathcal K\big(\cosh\rho\Gamma_{\underline{r}}-\sinh\rho\sinh r\Gamma_{\underline{\rho}}\big)-i\sinh\rho\cosh r \Gamma_{\underline{\rho r}}\GammaAdS~,
 \\
 \mathcal R_A[\Gamma_{\underline{\rho r}}]&=
 -i\mathcal K\big(\sinh\rho\Gamma_{\underline{r}}-\cosh\rho\sinh r\Gamma_{\underline{\rho}}\big)\GammaAdS
 +\cosh\rho\cosh r\Gamma_{\underline{\rho r}}~,
\end{align}
\end{subequations}
where the matrix $\mathcal K$ encodes the entire dependence on the AdS$_3$ directions and is given by
\begin{align}
 \mathcal K&=\big[e^{-x}+e^x(y^2-t^2)\big]P_{x+}+e^xP_{x-}-ie^x\big(t\Gamma_{\underline{t}}+y\Gamma_{\underline{y}}\big)\GammaAdS~.
\end{align}
The projectors $P_{x\pm}$ were defined above (\ref{eqn:AdS-R-matrix}).
We now use that these structures act on $\epsilon_0$, which satisfies $\Gamma_{11}\epsilon_0=\epsilon_0$ and 
the massive projection condition (\ref{eqn:massive-projector}).
So we have $\GammaAdS\epsilon_0=-\GammaS \epsilon_0$ and 
$\Gamma_{\underline{\rho r}}\epsilon_0=-\hat \Gamma_p\epsilon_0$,
with $\hat\Gamma_p=\frac{\lambda}{|c|}\Gamma_p$ such that $\hat\Gamma_p^2=-\mathds{1}$.
This yields
\begin{subequations}
\begin{align}
 \mathcal R_A[\Gamma_{\underline{\rho}}]\epsilon_0&=
 \cosh r\mathcal K\Gamma_{\underline{\rho}}\epsilon_0+i\sinh r \hat\Gamma_p\GammaS\epsilon_0~,
 \\
 \mathcal R_A[\Gamma_{\underline{r}}]\epsilon_0&=
 -\big(\cosh\rho \hat \Gamma_p + \sinh\rho\sinh r\,\mathds{1}\big)\mathcal K\Gamma_{\underline{\rho}}\epsilon_0
 -i\sinh\rho\cosh r\,\hat\Gamma_p\GammaS\epsilon_0~,
 \\
 \mathcal R_A[\Gamma_{\underline{\rho r}}]\epsilon_0&=
 i\big(\sinh\rho \hat\Gamma_p+\cosh\rho\sinh r\,\mathds{1}\big)\GammaS\mathcal K\Gamma_{\underline{\rho}}\epsilon_0
 -\cosh\rho\cosh r\,\hat\Gamma_p\epsilon_0~.
\end{align}
\end{subequations}
We also used that $\hat\Gamma_p$ has an even number of S$^5$ $\Gamma$-matrices and commutes with AdS $\Gamma$-matrices,
and that $\mathcal K$ commutes with S$^5$ $\Gamma$-matrices.
With these identities, eq.~(\ref{eqn:kappa-non-linear-RA}) becomes
\begin{align}
\mathcal Q_{\mathcal K}\,\mathcal K\Gamma_{\underline{\rho}}\epsilon_0 
+\mathcal Q_{\mathds{1}}\epsilon_0&=h\epsilon_0~,
\end{align}
where
\begin{subequations}
\begin{align}
\begin{split}
 \mathcal Q_{\mathcal K}&=
 \cosh r\mathcal R_S[\mathcal N^{\rho}]
 -\cosh^2\!\rho\,\mathcal R_S[\mathcal N^r]\big(\hat\Gamma_p+\tanh\rho\sinh r\mathds{1}\big)
 \\ &\hphantom{=}\,
 +i\cosh^2\!\rho\,\mathcal R_S[\mathcal N^{\rho r}]\big(\tanh\rho \hat\Gamma_p+\sinh r\mathds{1}\big)\GammaS~,
\end{split}
 \\
\begin{split}
 \mathcal Q_{\mathds{1}}&=
 \mathcal R_S[\mathcal N^{\mathds{1}}]+i \sinh r\mathcal R_S[\mathcal N^\rho]\hat\Gamma_p \GammaS
 -i\sinh\rho\cosh\rho\cosh r\mathcal R_S[\mathcal N^r]\hat\Gamma_p\GammaS
 \\&\hphantom{=}\,
 -\cosh^2\!\rho\,\cosh r\mathcal R_S[\mathcal N^{\rho r}]\hat\Gamma_p~.
\end{split}
\end{align}
\end{subequations}
We have implemented all the projectors and chirality conditions we have at our disposal to reduce the AdS Clifford algebra structures to just one, 
and we can hence now formulate the conditions for $\kappa$-symmetry as two operator equations which have to be satisfied simultaneously. 
This leads to the conditions in eq.~(\ref{eqn:kappa-non-linear-B}).

\bibliography{susyD3D5}
\end{document}